\documentclass[twocolumn,aps,prx,showpacs,preprintnumbers,amsmath,amssymb,superscriptaddress,longbibliography]{revtex4-2}

\usepackage{graphicx}    
\usepackage{latexsym}    %
\usepackage{color}       %
\usepackage{dcolumn}     
\usepackage{bm}          
\usepackage{amssymb}     
\usepackage{hyperref}    
\expandafter\let\csname equation*\endcsname\relax
\expandafter\let\csname endequation*\endcsname\relax
\usepackage{amsmath}
\usepackage{mathtools}          

\usepackage{graphics}
\usepackage{epsfig}
\usepackage{subfigure}
\usepackage{nicefrac}
\usepackage{verbatim}
\usepackage{enumerate}
\usepackage{footnote}
\usepackage{booktabs}
\usepackage{siunitx}
\usepackage[noline]{algorithm2e}

\def\url#1{}
\def\urlprefix{}

\begin{document}

\title{Superstatistical wind fields from point-wise atmospheric turbulence measurements}

\author{J.~Friedrich}
\affiliation{ForWind, Institute of Physics, University of Oldenburg
K\"upkersweg 70, D-26129 Oldenburg, Germany}
\author{D.~Moreno}
\affiliation{ForWind, Institute of Physics, University of Oldenburg
K\"upkersweg 70, D-26129 Oldenburg, Germany}
\author{M.~Sinhuber}
\affiliation{ForWind, Institute of Physics, University of Oldenburg
K\"upkersweg 70, D-26129 Oldenburg, Germany}
\author{M.~W\"achter}
\affiliation{ForWind, Institute of Physics, University of Oldenburg
K\"upkersweg 70, D-26129 Oldenburg, Germany}
\author{J.~Peinke}
\affiliation{ForWind, Institute of Physics, University of Oldenburg
K\"upkersweg 70, D-26129 Oldenburg, Germany}
\date{\today}

\begin{abstract}
Accurate models of turbulent wind fields have become increasingly important in the atmospheric sciences, e.g., for the  determination of
spatiotemporal correlations in wind parks, the estimation of individual loads on turbine rotor and blades, or the modeling of particle-turbulence interaction in atmospheric clouds or pollutant distributions in urban settings. Due to the difficult task of resolving the fields across a broad range of scales, one oftentimes has to invoke stochastic wind field models that fulfill specific, empirically observed, properties.
Whereas commonly used Gaussian random field models solely control second-order statistics (i.e., velocity correlation tensors or kinetic energy spectra), we explicitly show that our extended model emulates the effects of higher-order statistics as well. Most importantly, the empirically observed phenomenon of small-scale intermittency, which can be regarded as one of the key features of atmospheric turbulent flows, is reproduced with a very high level of accuracy and at considerably low computational cost. Our method is based on a multipoint statistical description of turbulent velocity fields that consists of a superposition of multivariate Gaussian statistics with fluctuating covariances. We propose a new and efficient sampling algorithm for this Gaussian scale mixture and
demonstrate how such ``superstatistical'' wind fields can be constrained on a certain number of real-world measurement data points from a meteorological mast array. 
\end{abstract}

\maketitle

\section{Introduction}
\label{sec:intro}
Atmospheric turbulence is one of the key drivers of atmospheric processes, and strongly affects properties such as mixing or energy and momentum transfer in the atmospheric boundary layer~\cite{wyngaard1992atmospheric,sutton2020atmospheric}. Since the early works by Richardson~\cite{richardson1921some,richardson2007weather}, Monin~\cite{Monin_1958}, and Oboukhov~\cite{Oboukhov1962}, atmospheric turbulence research has been carried out by joint efforts between experimental, theoretical, as well as numerical approaches. Nonetheless, our understanding of the basic physical processes underlying the phenomenon of atmospheric turbulence, e.g., the empirically confirmed higher-order statistics of small-scale wind field fluctuations~\cite{morales2012characterization,wachter2012turbulent}, is rather limited.
Therefore, the treatment of the long-standing problem
of atmospheric turbulent flows has to be addressed by a combination of state-of-the-art
experimental measurement campaigns~\cite{mikkelsen2014lidar,dfwind,neuhaus2020generation} and methods from nonequilibrium statistical physics~\cite{Cardy_2008,BECK2003267,metzler2020superstatistics}.

The need for more accurate models of atmospheric turbulence can best be illustrated in the context of the wind energy sciences: Due to the ever-increasing size of wind turbines, with recent generations potentially exceeding the height of the atmospheric surface layer, new challenges for turbine design and operating conditions arise~\cite{Veers:2019aa,Meneveau_2019}. As it is computationally expensive to resolve the broad range of scales from rotor diameter to dissipative scales \emph{and} to account for the unsteadiness of atmospheric turbulence within direct numerical simulations (DNS) of the governing fluid dynamical equations, one oftentimes has to resort to stochastic inflow turbulence models for the assessment of turbine loads and power output. On the other hand, in large eddy simulations (LES), which are computationally less demanding than DNS, the use of certain subgrid models does not preserve turbulence properties down to an arbitrarily small scale.
Furthermore, a considerable amount
of the computational time has to be invested into the build-up of flow profiles with the desired turbulence characteristics (e.g., mean velocity profile or turbulence intensity). Therefore, spectral stochastic models such as the ones proposed by Mann and Veers~\cite{MANN1998269,veers1984modeling} play an important role for the design process of wind turbines, which is also reflected by the guidelines established by the International Electrotechnical Commission~\cite{international2005wind}.  Whereas such spectral models capture spatiotemporal correlations as well as shear, they do not account for the non-Gaussian/intermittent features of atmospheric turbulence, and therefore potentially bias subsequent load calculations~\cite{muecke,hannesdottir2019extreme,Gontier_2007}.
On the other hand, non-Gaussian models, e.g., the Continuous Time Random Walk model~\cite{kleinhans2006simulation,muecke,Yassin_2021}
control intermittency properties of a turbulent time series (by a method referred to as subordination of a stochastic process~\cite{fogedby1994langevin,Eule_2009}), but not of a full three-dimensional field. As turbine load calculations are strongly affected by  three-dimensional wind field structures, the assessment of non-Gaussian wind fields statistics on load time series and material fatigue remains inconclusive~\cite{kleinhans2008stochastic}.
Here, we present such a model for the generation of a fully three-dimensional wind field that can be apprehended as an extension of the well-known Mann model for inflow turbulence.

The purpose of the present article is twofold: First, we aim at generalizing Gaussian wind field models~\cite{veers1984modeling,MANN1998269} to a novel class of random fields, which will be henceforth referred to as \emph{superstatistical random fields}, whose non-Gaussian properties are precisely controllable. Second, we propose a new method that constrains these random fields on sparse, point-wise atmospheric turbulence measurements; in our case from propeller anemometers in a meteorological mast array. Our approach thus addresses the problem of incomplete measurements which arises for instance in laser-based Doppler anemometer measurements~\cite{Beck_2017,amt-15-1355-2022} or due to large surface areas covered by aerial measurements of the wakes of large wind park clusters~\cite{x_wakes}. It has to be stressed that although the present work discusses the reconstruction of a wind field in front of a wind turbine, our methodology is applicable to a broad range of problems in atmospherics physics and beyond. Typical examples include the pollutant distribution in urban street canyons~\cite{buccolieri2015breathability,salizzoni2011turbulent}, the problem of urban heat island formation~\cite{voelkel2017towards}, the further development of sub-grid models~\cite{cassiani2010stochastic}, and, in a different context, the reconstruction of spatial fields in urban systems~\cite{epb}. Nonetheless, an extension of the here-proposed methodology requires additional knowledge, e.g., on the joint temperature-velocity field statistics, on the effects of atmospheric stability or on variable pressure gradients due to complex geometries.

This paper is organized as follows: Sec.~\ref{sec:charac} characterizes the atmospheric turbulence measurements that will serve to illustrate our methodology in Sec.~\ref{sec:superstat} and Sec.~\ref{sec:interpol}. Future model improvements as well as concluding remarks are given in Sec.~\ref{sec:improve}.
\section{Characterization of the atmospheric turbulence measurements}
\label{sec:charac}
In this section, we give a brief overview of the atmospheric wind field measurement campaign which will be the basis for the wind field reconstructions in Sec.~\ref{sec:interpol}. The GROWIAN (German for ``Gro\ss e Windenergieanlage'') measurement campaign
at Kaiser-Wilhelm-Koog, Germany, which was carried out intermittently in between January 1984 and February 1987, collected horizontal wind speed time series, measured by 16 propeller anemometers with a sampling frequency of 2.5\;Hz~\cite{koerber19883,dwd}. The propeller anemometers were
mounted on two met masts in front of a 3MW wind turbine, which at that time was the largest wind turbine worldwide. The turbine was a two-bladed ``lee-runner'' (its rotor faced downstream) and had a hub height of 102\;m.
The objective of the measurements was to gain insights into correlations between spatiotemporal wind structures and rotor loads~\cite{winter2016modellierung}.
\begin{figure}[h!]
\centering
\hspace{0.1cm}
\includegraphics[width=0.495 \textwidth]{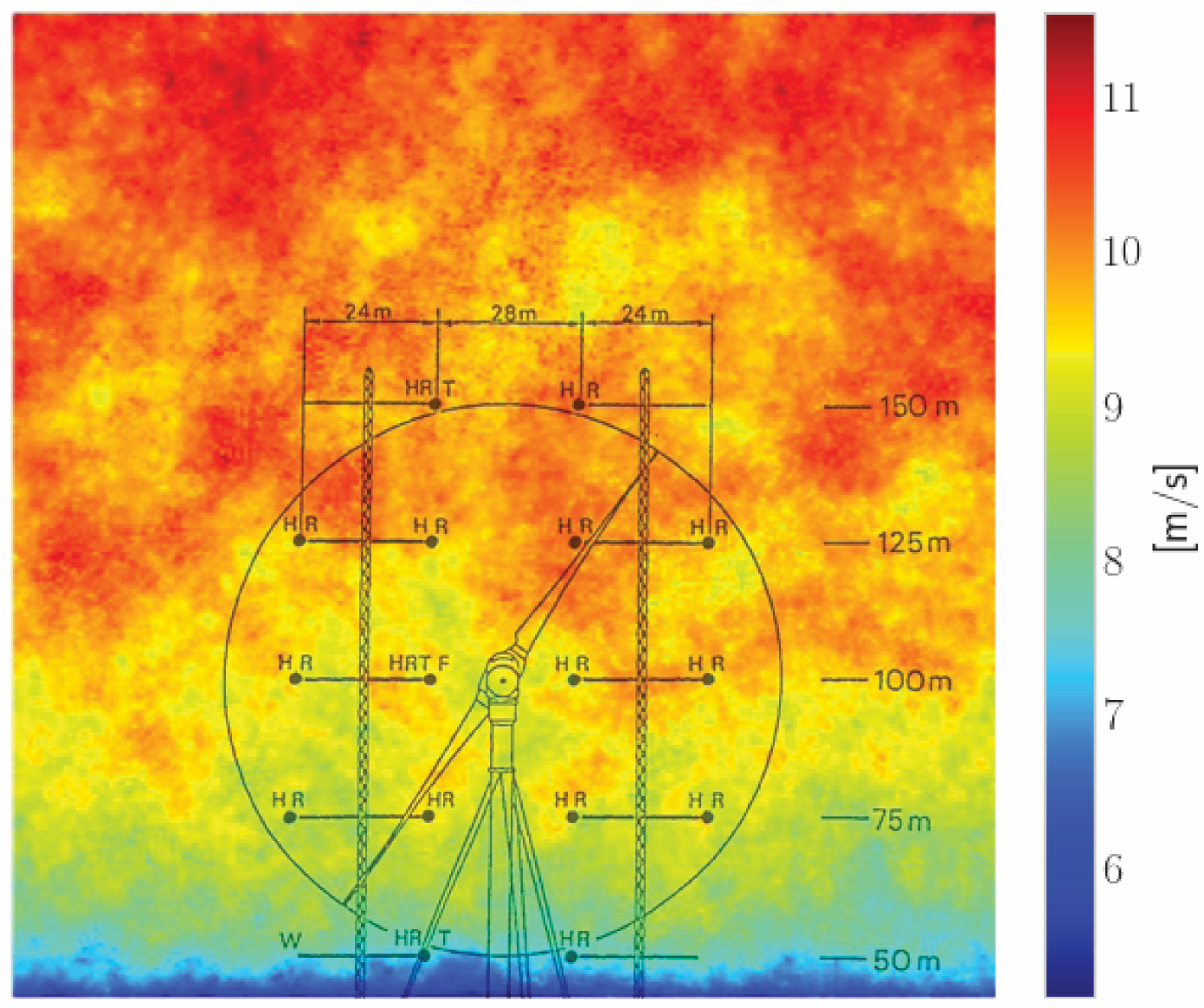}
\caption{Setup of the GROWIAN measurements at Kaiser-Wilhelm-Koog, Germany~\cite{koerber19883}. Full circles indicate the 16 propeller anemometers which were mounted on two meteorological masts. The 16 point-wise measurements are now used for reconstructing a full three-dimensional velocity field. A slice through the reconstructed velocity field is shown exemplarily (see Sec.~\ref{sec:superstat} and Sec.~\ref{sec:interpol} for further details on the reconstruction).}
\label{fig:growian_array}
\end{figure}

\begin{figure*}[ht]
\centering
\includegraphics[width=0.46 \textwidth]{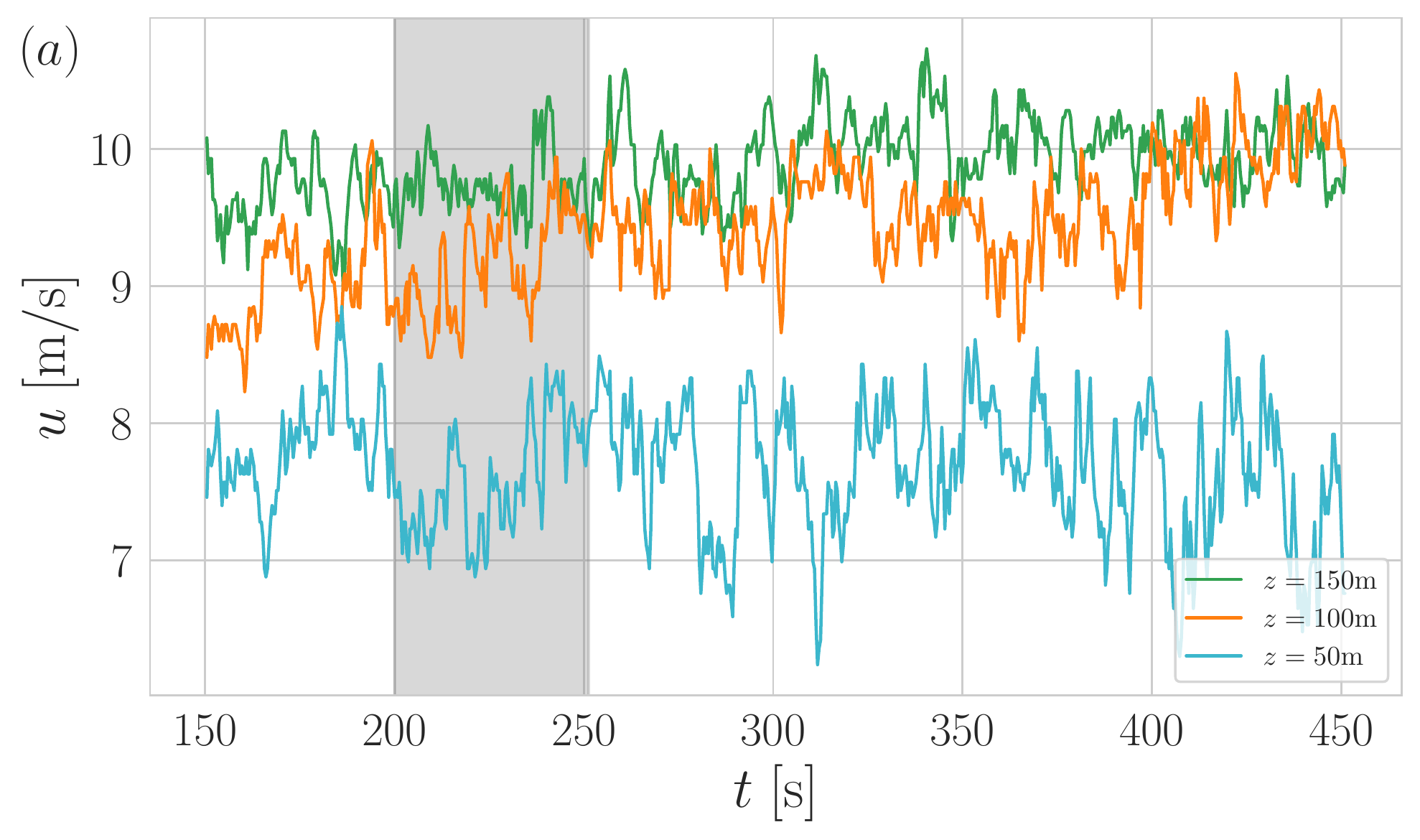}
\includegraphics[width=0.465 \textwidth]{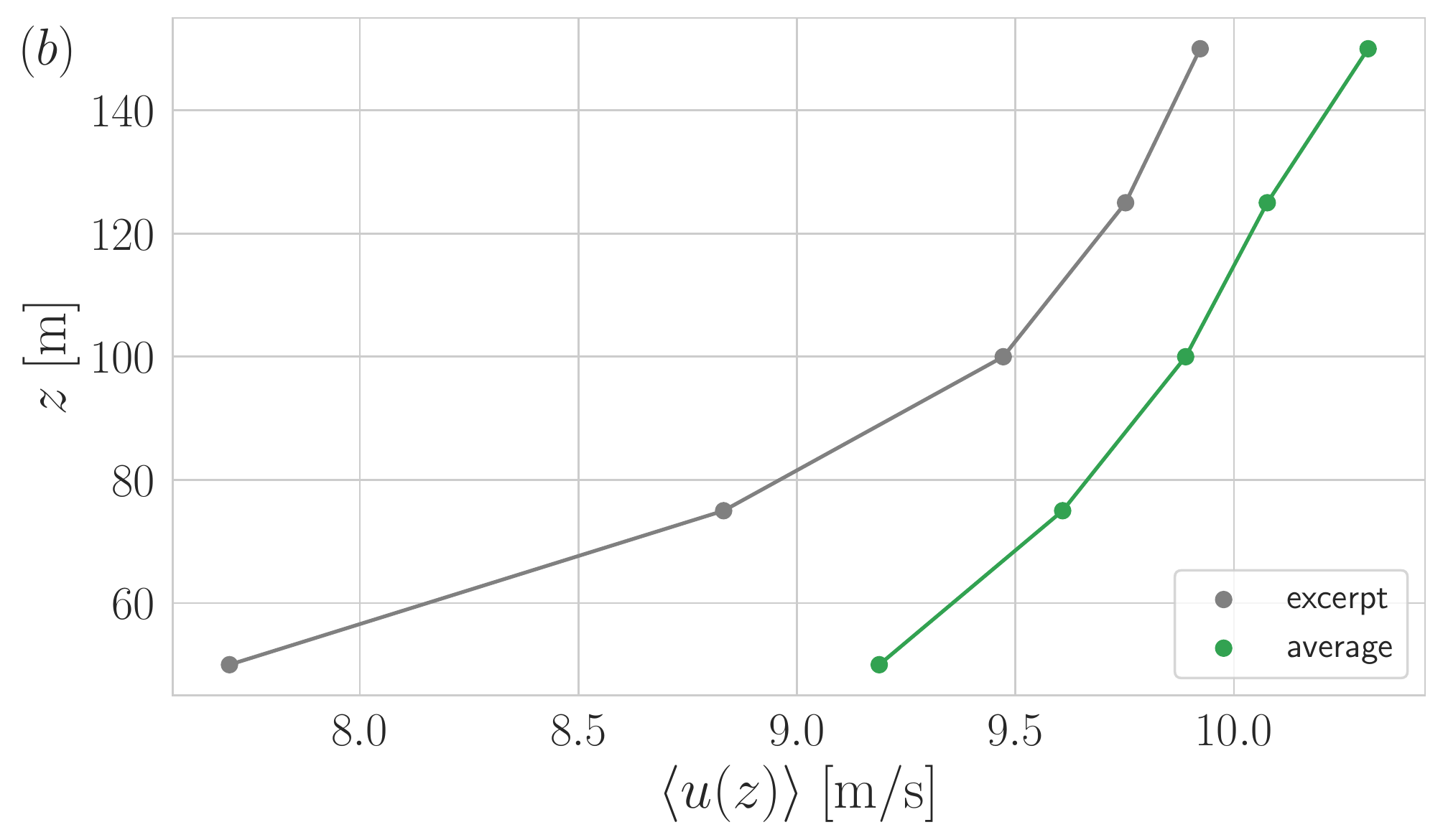}\\
\hspace{0.2cm} \includegraphics[width=0.45 \textwidth]{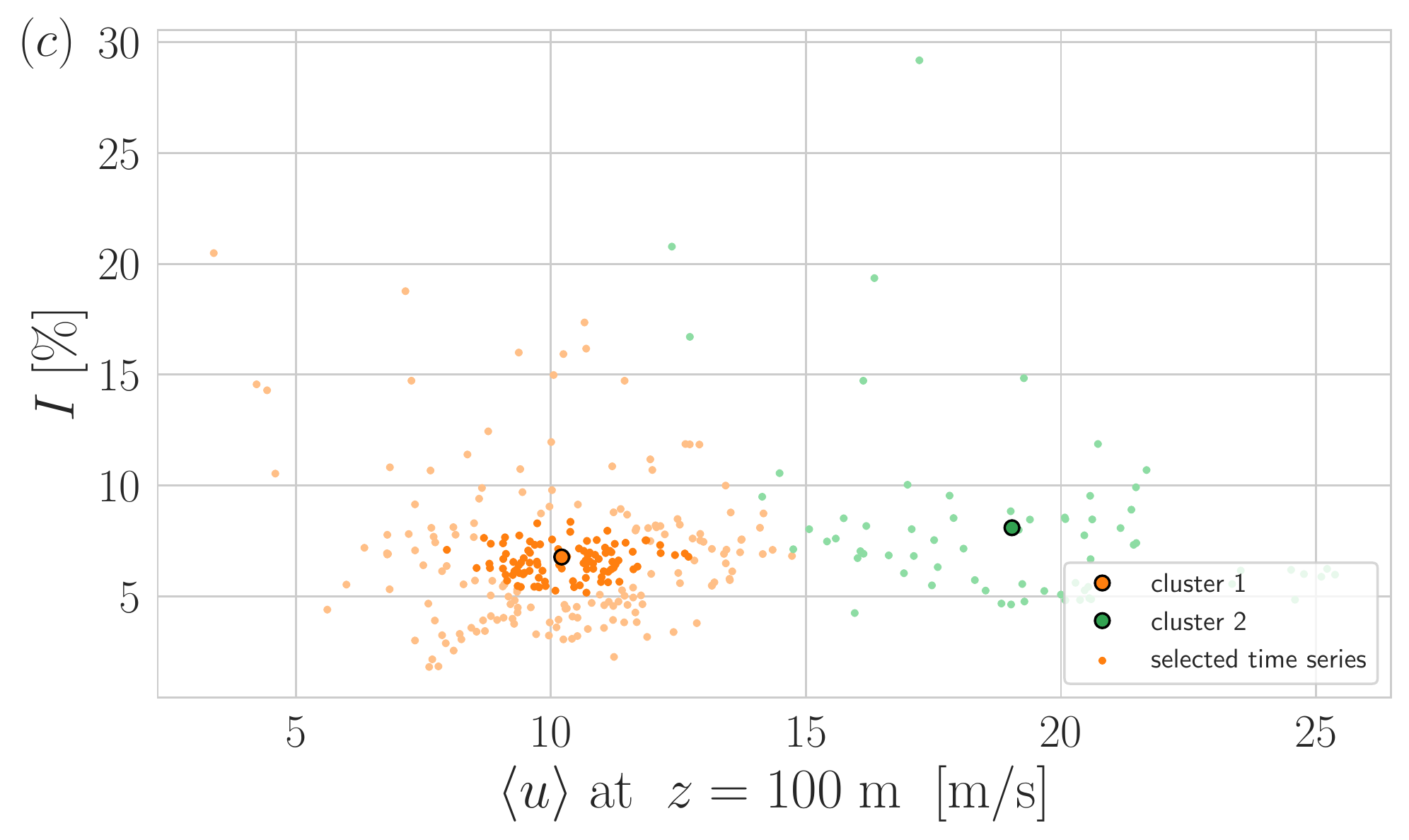}
\hspace{0.1cm}
\includegraphics[width=0.48 \textwidth]{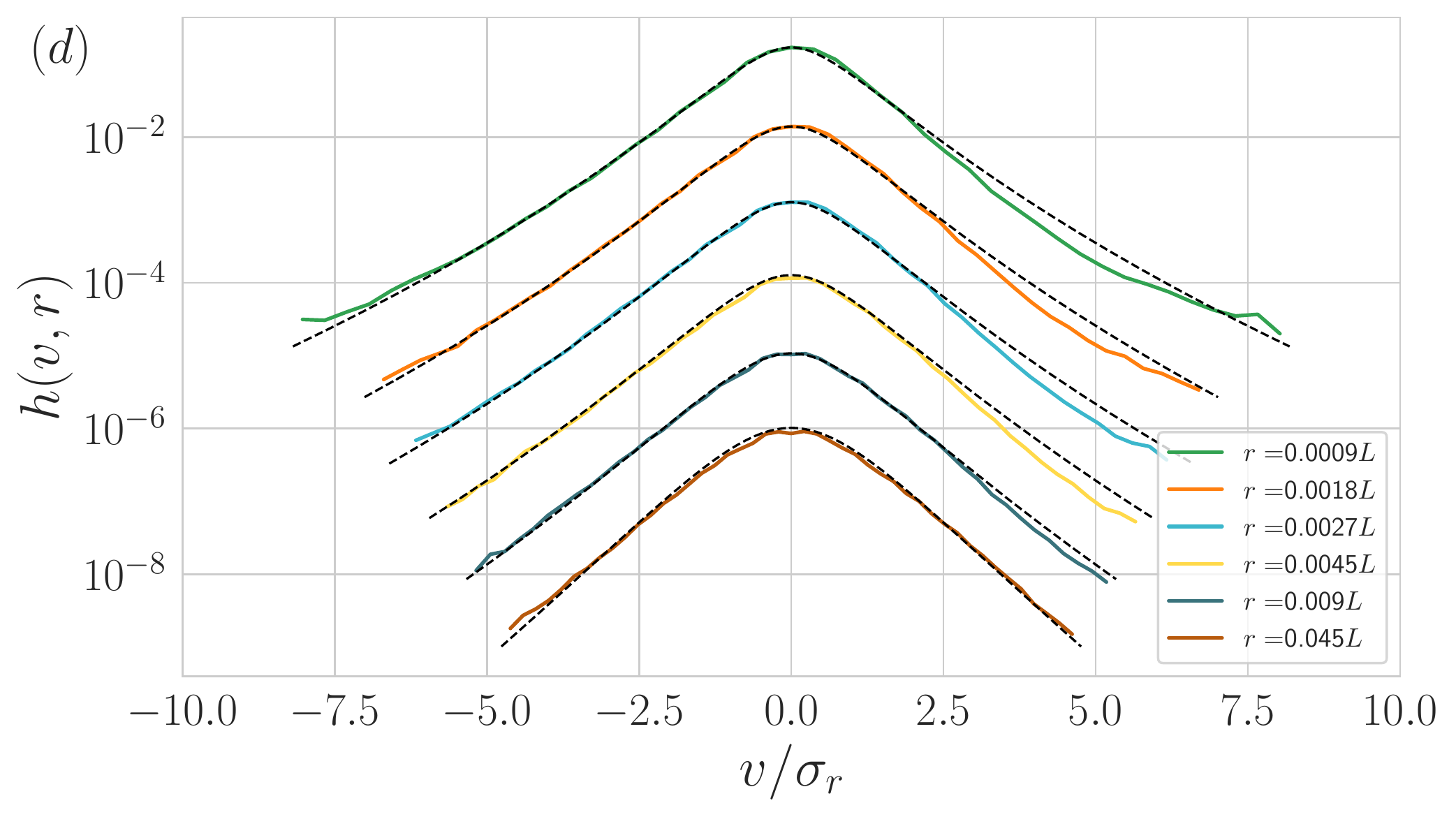}
\caption{(a) Typical excerpts of wind speed time series measured at one of the two met masts at the GROWIAN site at three different heights, 50, 100, and 150$\;$m (see Sec.~\ref{sec:charac} for further details on this measurement campaign). The hub height of the installed wind turbine was at $102$\;m. The gray-shaded area indicates the part of the time series that will be used to reconstruct a fully three-dimensional field in Sec.~\ref{sec:interpol}. (b) Vertical velocity profiles averaged over $x$-$y$-coordinates of the wind speed extracts similar to the gray-shaded ones in (a). The green line shows a horizontal average over all 100 time series that were part of the GROWIAN measurement campaign. (c) Turbulence intensity $I$ against mean wind speed $\langle u \rangle$ at height  $z=$100 m of all 334 GROWIAN measurements (green and orange points). Two clusters can be distinguished and 100 time series in the vicinity of cluster 1 (orange) have been selected (dark orange) for further statistical analysis.
(d) Velocity increment probability distribution averaged over all 100 time series of the GROWIAN measurements (dark orange points in (c)). The scale separations $r$ have been calculated using Taylor's hypothesis and are given in multiples of the model integral length scale $L$ (see Sec.~\ref{sec:charac} for further description). Dashed lines correspond to fits using the K62 model of homogeneous isotropic turbulence. Here, only the left part of the PDFs have been fitted as the PDFs exhibit a considerable asymmetry (the K62 model as stated here possesses no skewness) which becomes more pronounced at small scales. The precise reason for this asymmetry remains inconclusive, and it is unclear whether this is an intrinsic feature of the atmospheric turbulence (skewness is also predicted for homogeneous isotropic turbulence) or whether this is a measurement artifact, e.g., due to propeller inertia or increased response time of propeller alignment with changing wind direction~\cite{horst1973corrections}.}
\label{fig:growian_data}
\end{figure*}

In the context of wind field reconstructions, the measurements are particularly valuable since they cover an area spanning $76\times100$\;m$^2$ as depicted in Fig~\ref{fig:growian_array}. Fig.~\ref{fig:growian_data}(a) shows three exemplary time series $u(t)$ at heights 50, 100, and 150\;m. The gray-shaded area indicates the part of the 16 time series that will be used for reconstructing a full spatiotemporal wind field in Sec.~\ref{sec:interpol} and corresponds to $51.6$\;s. We choose the latter extract of the time series due to its clear
mean vertical velocity profile depicted in Fig.~\ref{fig:growian_data}(b), which has been obtained by averaging the 16 time series in horizontal direction. It must be stressed that the latter choice is rather arbitrary and that other parts of the time
series might exhibit entirely different vertical profiles (e.g., inverted or negative wind profiles for smaller time intervals).

In the following, we are interested in a more detailed statistical analysis of GROWIAN measurements and
extracting wind field parameters, which will be needed for our modeling endeavors in Sec.~\ref{sec:superstat}. To this end, we perform a pre-selection procedure of
time series from the entire available GROWIAN measurements which is motivated by the one proposed by M\"ucke et al.~\cite{muecke}. The first step of the procedure consists in calculating the mean and turbulence intensity of each of the 20 min time series at a given height (here, we choose a height of 100 m since it is close to the hub height of the original wind turbine at 102 m). Fig.~\ref{fig:growian_data}(c) depicts the turbulence intensity $I$ against mean $\langle u \rangle$ at $z=100$ m of all 344 available data sets (green and orange points).  
By the use of a simple K-means algorithm we can distinguish between two clusters, namely a cluster with moderate turbulence intensity and mean wind speed (cluster 1, orange), and a cluster with strong wind speed but moderate turbulence intensity (cluster 2, green). In the following, we select 100 nearest neighbor time series of cluster 1 (dark orange points) as it 
is more significant than cluster 2 and thus offers sufficient statistics for our purposes (the choice of exactly 100 time series is motivated by the Sandia report for subsequent load estimations~\cite{veers1988three}). 
Furthermore, as outlined in the original project report~\cite{koerber19883}, cluster 1 
corresponds to stable atmospheric conditions whereas cluster 2 can be mainly attributed to unstable stratification with strong wind gusts. In order to test for the stationarity of these time series, we simply split each 20 min series into half and verify whether 10 min mean values and turbulence intensities deviate significantly or not (relative errors for mean and turbulence intensity are 2.01\% and 9.23\% for the 100 selected time series, respectively). Hence, by means of this selection procedure, we try to ensure that the data sets correspond to a stationary wind time series.

In the context of the further statistical analysis, we must emphasize that  several hypothesis from the theory of homogeneous isotropic turbulence~\cite{frisch:1995} will be invoked in a \emph{local sense}, i.e., at small scales where the turbulent fluctuations ``forget'' about the presence of large-scale atmospheric motions or circulations. 
We first use Taylor's hypothesis in order to convert the temporal time series into a spatial field~\cite{Taylor1935}. Here, we assume that the meteorological mast array in Fig.~\ref{fig:growian_array} spans the $y$-$z$-plane, whereas the $x$-direction is obtained from $x=\langle u_x \rangle t$, where
$\langle u_x \rangle$ denotes the mean velocity in $x$-direction $u_x$, which can be calculated from wind speed and wind direction measurement~\cite{muecke}. Furthermore,
the theory of homogeneous isotropic turbulence
explicitly excludes the presence of a mean velocity field, which contradicts the observed mean vertical velocity profiles in Fig.~\ref{fig:growian_data}(b). To assess the small-scale fluctuations, we therefore subtract the mean vertical velocity profile from the measurements.
The corresponding probability density function (PDF) of longitudinal velocity increments $v(r)=\left[ \mathbf{u}(\mathbf{x}+\mathbf{r})-\mathbf{u}(\mathbf{x})\right]\cdot \frac{\mathbf{r}}{r}$
at different scales $r$ are depicted in Fig.~\ref{fig:growian_data}(d). 
Deviations from Gaussianity in Fig.~\ref{fig:growian_data}(d) are visible at all observable scales $r$. Even at large scales $r$, the PDFs do not approach a Gaussian distribution, which contrasts observations from homogeneous isotropic turbulence~\cite{Friedrich1997}. 
This particularity of atmospheric turbulent flows can be attributed to non-stationary features of atmospheric flows and has been observed in numerous atmospheric turbulence studies~\cite{morales2012characterization,boettcher2003statistics}. In more detail, if the 
velocity field would exhibit stationary features at large scales $r$,  the PDFs $h(v,r)$ would decay into a single-point quantity which is close to Gaussian.
Nonetheless, here, we are solely interested in reproducing the evolution
of small-scale atmospheric turbulent fluctuations
by a model of homogeneous isotropic turbulence, the Kolmogorov-Oboukhov K62 model (see~Sec.~\ref{sec:joint} for further discussion). The dashed lines in Fig.~\ref{fig:growian_data}(d) indicate the K62 model predictions with fit parameters $H=0.338$, $\mu =0.243$, and $L=10.75\;\textrm{km}$ (Hurst exponent $H$, intermittency coefficient $\mu$, and integral length scales $L$, which will be further specified in~Sec.~\ref{sec:joint}) and reproduce the evolution of small-scale fluctuations fairly well. 
We must emphasize that the here-determined model length scale $L$ does not correspond to a characteristic turbulence length of the atmospheric boundary layer, but solely represents a large scale quantity which is consistent with the evolution of the PDF of small-scale fluctuations as depicted in Fig.~\ref{fig:growian_data}(d). In particular, the rather large value of $L$ can be attributed to long-range correlations of velocity field fluctuations in the analyzed data set.
Nonetheless, as we are mostly interested in reproducing the effects of small-scale fluctuations, latter parameters can now be used for reconstructing highly-resolved velocity fields as described in the following Sections.

\section{Superstatistical random fields}
\label{sec:superstat}
In this section, we outline a method for the synthesis of a random field $\mathbf{u}(\mathbf{x})$ which possesses
multifractal properties and, hence, is able to reproduce the non-Gaussian features of atmospheric turbulence measurements (as manifested, e.g, in Fig.~\ref{fig:growian_data}(d)). Similar
models of synthetic turbulence have been proposed in terms of  multiplicative cascade models ~\cite{juneja1994synthetic,malara2016fast,rosales2008anomalous,lovejoy1986scale} or random multifractal walks~\cite{bacry2001multifractal,chevillard2010stochastic}. Nonetheless, due to their complexity, difficulties in numerical implementations, as well as their inability to incorporate real point-wise measurements, latter models are barely used in the field of atmospheric turbulence. The here-proposed modeling approach therefore, addresses these issues in a systematic and practical manner by generalizing the well-known Mann model of inflow turbulence~\cite{MANN1998269}  to a non-Gaussian or superstatistical random field.

\begin{figure}[h!]
\centering
\includegraphics[width=0.47 \textwidth]{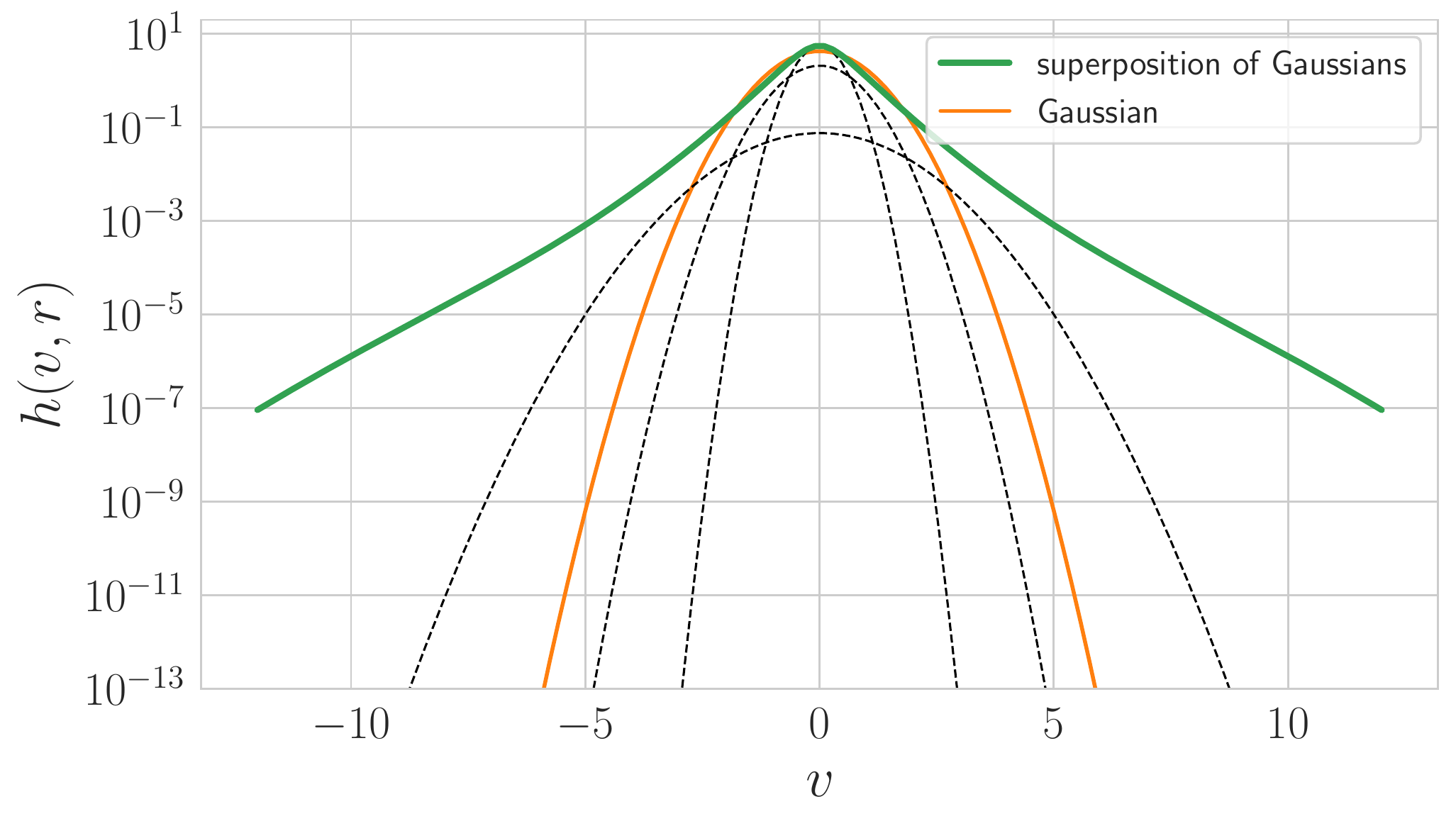}
\caption{(a) Schematic depiction the velocity increment statistics of the K62 model of turbulence. The green curve exhibits heavy tails due to a superposition of several Gaussian distributions (the dashed curves show three examples) with fluctuating variances, see also Eq. (\ref{eq:k62}). For comparison, the orange curve shows a Gaussian with the same standard deviation as the green, heavy-tailed, curve.}
\label{fig:schematic}
\end{figure}
In the following, we first derive a joint multipoint statistics for such a superstatistical random field
(see also~\cite{Friedrich_2021}) and prove that its velocity increment statistics (a two-point quantity) is consistent with the Kolmogorov-Oboukhov (K62) model of turbulence. Hence, in a first iteration, we approximate the atmospheric wind field as a homogeneous and isotropic random field. Therefore, empirically observed features such as wind shear or atmospheric stability are not intrinsically contained, but enter externally by constraining on measurements (see Sec.~\ref{sec:interpol}). Moreover, by imposing further conditions, such as the incompressibility of the velocity field, we obtain additional restrictions, in particular a relation between the statistics of longitudinal and transverse velocity increment. 

Hence, the model assumptions can be roughly summarized as follows:
\begin{enumerate}
\item the turbulent flow is rotationally invariant (assumption of isotropy)
\item the turbulent flow is invariant under translations (assumption of homogeneity)
\item the turbulent flow is incompressible (i.e., velocities are small compared to the speed of sound - a valid assumption in the Earth's atmosphere.)
\item assumptions 1.-4. only apply in a \emph{local sense}, large-scale flow patterns are influenced by anisotropies and shear; these effects are taken into account by constraining the resulting superstatistical random field on existing turbulence measurements
\end{enumerate}
The following model derivation is organized as follows: Sec.~\ref{sec:joint} derives a joint multipoint statistics which serves as a starting point for the sampling algorithm of superstatistical random fields described in Sec.~\ref{sec:sample}. Sec.~\ref{sec:interpol} details how real-world measurement points can be embedded in such superstatistical random fields while preserving the intermittency properties at small scales.
\subsection{Joint multipoint statistics of superstatistical random fields}
\label{sec:joint}
The central idea of our approach can be traced back to the works of Kolmogorov~\cite{Kolmogorov1962} and Oboukhov~\cite{Oboukhov1962} who devised a phenomenological model for the longitudinal velocity increments $v(r)=\left[ \mathbf{u}(\mathbf{x}+\mathbf{r})-\mathbf{u}(\mathbf{x})\right]\cdot \frac{\mathbf{r}}{r}$.
As mentioned in the previous section, the statistics of $v(r)$ in atmospheric turbulence is dominated by the occurrence of extreme
events at small length scales $r$ (heavy tails of the probability density function) and deviates significantly from Gaussian statistics. In the K62 model,
non-Gaussian behavior emerges due to a superposition of Gaussian distributions with varying variances which is schematically depicted in Fig.~\ref{fig:schematic}.
The statistics of $v(r)$ (green curve) has been obtained by superposing several Gaussians (dashed lines), which results in pronounced tails, or in other words,
an increased probability for the occurrence of extreme fluctuations of $v(r)$.

In mathematical terms, the probability density function (PDF)
of the longitudinal velocity increments $v(r)$ in the K62 model
\begin{equation}
  h(v,r)= \int_0^\infty \textrm{d}\xi g(\xi)\frac{1}{\sqrt{2 \pi}\sigma_{\xi}(r)} e^{-\frac{v^2}{2\sigma_{\xi}^2(r)}}\;,
  \label{eq:k62}
\end{equation}
is calculated as a superposition of Gaussian distributions whose variances
\begin{equation}
  \sigma_{\xi}^2(r) = \sigma^2
  \xi^{2H\sqrt{A+\mu \ln_+ \frac{x^{\succ}}{r+x^{\prec}}}}\left(\frac{r+x^{\prec}}{x^{\succ}}\right)^{H\mu}r^{2H}\;,
  \label{eq:var_k62}
\end{equation}
fluctuate with respect to the parameter $\xi$, which follows a lognormal distribution (we also refer the reader to~\cite{Friedrich_2021})
\begin{equation}
  g(\xi) = \frac{1}{\sqrt{2 \pi}\xi}e^{-\frac{[\ln \xi]^2}{2}}\;.
  \label{eq:lognormal}
\end{equation}
Moreover, the intermittency coefficient $\mu$ determines deviations from a purely self-similar distribution which is characterized by the Hurst exponent $H$
($H=1/3$ in the K41 phenomenology~\cite{Kolmogorov1941}, where $x^{\prec}$ and $x^{\succ}$ are large- and small-scale cut-off, 
$\sigma$ is the intensity of fluctuations, and $A$ is a quantity which depends on the flow's large-scale structure~\cite{Kolmogorov1962}). The emergence of such nonequilibrium statistics as a consequence of the superposition of equilibrium statistics is sometimes referred to as \emph{superstatistics}~\cite{BECK2003267,beck2007statistics,CASTAING1990177,metzler2020superstatistics,Wilczek_2016,YAKHOT2006166}.

The K62 model can be considered as a rather realistic model of a turbulent flow; the empirically observed phenomenon of intermittency is captured in an accurate manner, which is also underlined by the fits in Fig~\ref{fig:growian_data}(d).

Nonetheless, a statistical model for the velocity increments $v(r)$ is not sufficient for generating a \emph{complete} random field $\mathbf{u}(\mathbf{x})$, as it solely relies on two-point statistical quantities~\cite{monin}. A generalization to a multipoint statistics that consist of a superposition of multivariate Gaussian statistics is however possible~\cite{Friedrich_2021} and yields the joint $n$-point PDF
\begin{align}\nonumber
 \lefteqn{f_n(\mathbf{u}_1,\mathbf{x}_1;\ldots;\mathbf{u}_n,\mathbf{x}_n)}\\ 
 =& \int_0^\infty \mathrm{d}\xi g(\xi)
 \frac{1}{\sqrt{(2\pi)^{3n} \det \sigma_\xi}}e^{-\frac{1}{2} u_{i,\alpha} \sigma_{\xi,i\alpha;j\beta}^{-1} u_{j,\beta}}\;,
 \label{eq:n_point_3d}
\end{align}
where Greek indices indicate spatial components $\alpha=1,2,3$, (summation over identical indices is implied). Furthermore, we introduce the covariance matrix $\sigma_{\xi,i\alpha;j\beta}=C_{\xi,\alpha \beta}(\mathbf{x}_i-\mathbf{x}_j)$ and its inverse $\sigma_{\xi,i\alpha;j\beta}^{-1}$ which are entirely determined by the correlation tensor $C_{\xi,\alpha \beta}(\mathbf{x}_i-\mathbf{x}_j)$ of the individual part of the Gaussian ensemble.
Similar to the variances of the velocity increment PDF (\ref{eq:var_k62}), these fluctuating covariances
introduce strong correlations between \emph{all}
spatial points in the flow. Furthermore,
the particular choice of $g(\xi)=\delta(\xi-\xi_0)$ in Eq. (\ref{eq:n_point_3d}), where $\delta$ denotes the Dirac delta distribution,
reduces the model to a Mann-type model and thus results in a  purely Gaussian random field $\mathbf{u}(\mathbf{x})$.

Before we further discuss how random velocity field realizations can be drawn from the distribution (\ref{eq:n_point_3d}), we further investigate statistical properties that are determined by the correlation tensor $C_{\xi,\alpha \beta}(\mathbf{x}_i-\mathbf{x}_j)$. As outlined in Sec.~\ref{sec:charac}, we assume that in a first-order approximation, we are dealing with homogeneous isotropic turbulence, whereas effects due to shear and atmospheric stability will be discussed in Sec.~\ref{sec:improve}. In the following, we thus determine the parametrization of the Gaussian scale mixture (mixing distribution $g(\xi)$ and covariance matrices $\sigma_{\xi}$ of the multivariate Gaussian distributions in Eq. (\ref{eq:n_point_3d})) by invoking a realistic phenomenological model of turbulence, i.e., the K62 model. Here, the empirically determined model parameters will enter the covariance matrices in a more intricate way than in the variances of the original K62 model (\ref{eq:var_k62}). In principle, it is also possible to directly infer the decomposition into Gaussian statistics suggested by Eq. (\ref{eq:n_point_3d}) on the basis of experimental measurements~\cite{Beck:2005vx}, e.g., in flows with complex geometry where the assumptions of homogeneous isotropic turbulence are not fulfilled.
Nonetheless, in the case of homogeneous isotropic turbulence, the correlation tensor $C_{\xi,\alpha \beta}(\mathbf{x}_i-\mathbf{x}_j)$ is solely a function of the separation
$\mathbf{x}_i-\mathbf{x}_j$, which implies that the entire joint $n$-point PDF is invariant with respect to translations~\cite{friedrich2020non}, $\mathbf{x}_i \rightarrow \mathbf{x}_i +\mathbf{X}$ and which thus ensures the homogeneity of the random field $\mathbf{u}(\mathbf{x})$. Furthermore, the assumption of isotropy (i.e., invariance under rotations) leads to the following ansatz for the correlation tensor~\cite{monin}
\begin{equation}
  C_{\xi,\alpha \beta}(\mathbf{r}) = \left(C_{\xi,rr}(r)-C_{\xi,tt}(r)\right)\frac{r_\alpha r_\beta}{r^2}+C_{\xi,tt}(r)\delta_{\alpha \beta}\;,
  \label{eq:corr_tensor}
\end{equation}
where $C_{\xi,rr}(r)$ and $C_{\xi,tt}(r)$ denote
the longitudinal and transverse correlation function, respectively, and $\delta_{ij}$ denotes the Kronecker delta.
In the following, we further investigate the implications of this ansatz for the statistics of the corresponding random field:

\emph{i.) incompressibility:} As shown in Appendix~\ref{app:incompressibility}, the incompressibility condition $\nabla \cdot \mathbf{u}(\mathbf{x})=0$, implies that the longitudinal and transverse correlation functions in Eq. (\ref{eq:corr_tensor}) are related by
 \begin{equation}
    \int_0^\infty \mathrm{d}\xi g(\xi)\left[ C_{\xi,tt}(r) - \frac{1}{2r}\frac{\partial}{\partial r}  \left(r^2 C_{\xi,rr}(r)\right)  \right]=0\;.
    \label{eq:karman}
  \end{equation}
As $g(\xi)$ is an arbitrary distribution (in our case the lognormal distribution in Eq.(\ref{eq:lognormal})), the integral can only vanish if the expression in the square brackets vanishes, hence, each of the individual parts of the Gaussian ensemble has to fulfill a von K\'arm\'an-Howarth relation~\cite{frisch:1995}
  \begin{equation}
    C_{\xi,tt}(r) = \frac{1}{2r}\frac{\partial}{\partial r}  \left(r^2 C_{\xi,rr}(r)\right),
    \label{eq:karman_naked}
  \end{equation}
which entails that the statistical properties of the random field (\ref{eq:n_point_3d}) are solely determined by the appropriate choice of either the longitudinal
or the transverse correlation functions. As we will specify the longitudinal statistics in \emph{iii.)} in more detail, here, we consider $C_{\xi,rr}(r)$
as variable and, consequently, $C_{\xi,tt}(r)$ as fixed by Eq. (\ref{eq:karman_naked}).
\par
\emph{ii.) single-point statistics:} The definition of the joint multipoint statistics in Eq. (\ref{eq:n_point_3d}) implies that the mean of the velocity field, $\langle \mathbf{u}(\mathbf{x}) \rangle$, vanishes. Therefore, the single-point statistics ($n=1$ in Eq. (\ref{eq:n_point_3d})) reduces to a centered Gaussian
\begin{equation}
  f_1(\mathbf{u}_1) = \frac{1}{\sqrt{2\pi}u_{rms} }e^{-\frac{u_1^2}{2 u_{rms}^2}}\;,
\end{equation}
where we used $C_{\xi,\alpha \beta}(\mathbf{r}=0)=u_{rms}^2\delta_{ij}$ with the root mean square velocity $u_{rms}= \sqrt{\langle \mathbf{u}^2 \rangle}$ that will be specified in terms of the model parameters later on. It should
be noted that the dependence of $f_1(\mathbf{u}_1,\mathbf{x}_1)$ on $\mathbf{x}_1$ vanishes as a consequence of the homogeneity condition.
%
\begin{figure*}[ht]
\includegraphics[width=0.65 \textwidth]{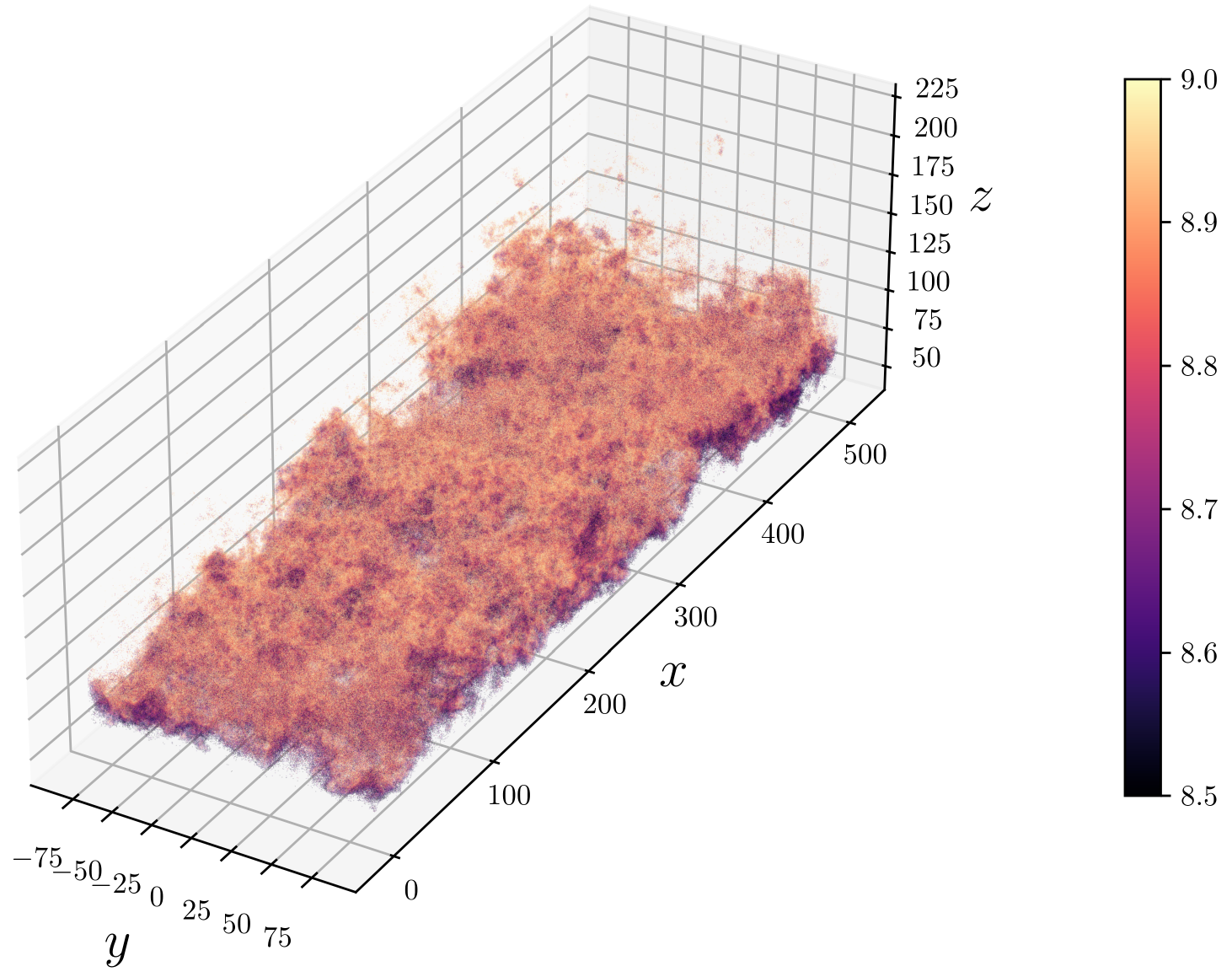}
\caption{Scatter plot of the reconstructed velocity field from the GROWIAN measurements (shaded gray area in Fig.~\ref{fig:growian_data}(a)). The points are chosen to lie within the interval $[8.5, 9] \frac{\textrm{m}}{\textrm{s}}$ and are color-coded with their corresponding values.}
\label{fig:scatter}
\end{figure*}

\emph{iii.) longitudinal velocity increment statistics \`a la K62:} As mentioned under \emph{i.)}, the entire statistical properties of the velocity field $\mathbf{u}(\mathbf{x})$ are fully determined by an appropriate choice of the longitudinal correlation function $C_{\xi,rr}(r)$. In the following, we impose an additional requirement, i.e., the reduction of the longitudinal increment statistics to the K62 model of turbulence as given by Eq. (\ref{eq:k62}).
Latter requirement parameterizes the longitudinal correlation function in terms of the model parameters (i.e., the intermittency coefficient $\mu$, the Hurst exponent $H$, the integral length scale $L$, as well as a large- and small-scale cutoff).
We first define the longitudinal structure functions as $S_{n,r}(r)= \langle v(r)^n \rangle$. As shown in Appendix~\ref{app:longi}, the longitudinal structure functions can be derived from the multipoint statistics as
\begin{equation}
  S_{n,r}(r)=   \int_0^\infty \mathrm{d}\xi g(\xi) (n-1)!! 2^{\frac{n}{2}} \left[S_{2\xi,r}(r) \right]^{\frac{n}{2}}\;,
  \label{eq:longi_struc}
\end{equation}
where $S_{2\xi,r}(r)=2C_{\xi,rr}(0)-2C_{\xi,rr}(r)$. Latter relation holds only for even $n$ and structure functions of odd order are zero as the model
does not account for skewness~\cite{Friedrich_2021}. In the following, we choose
\begin{widetext}
\begin{align}\nonumber
  &~C_{\xi,rr}(r)=-\frac{\sigma^2}{2} \rho_\xi(r)^{2H}  +\frac{\sigma^2 L^{2H}}{2}\Gamma(2H+1)\cosh \frac{\rho_\xi(r)}{L} \\
  ~&+ \frac{\sigma^2 \rho_\xi(r)^{2H+1} }{4L(2H+1)}  \left[
  e^{-\frac{\rho_\xi(r)}{L}} \mathcal{K}^H  \left(\frac{\rho_\xi(r)}{L} \right) -e^{\frac{\rho_\xi(r)}{L}} \mathcal{K}^H  \left(-\frac{\rho_\xi(r)}{L} \right) \right]\;,
  \label{eq:corr_lambda}
\end{align}
\end{widetext}
which is the correlation function of a fractional Ornstein-Uhlenbeck process~\cite{Mardoukhi_2020} with a re-parameterzied scale
\begin{equation}
  \rho_\xi(r)= \xi^{\sqrt{A+\mu \ln_+ \frac{x^{\succ}}{r+x^{\prec}}}}\left(\frac{r+x^{\prec}}{x^{\succ}}\right)^{\frac{\mu}{2}}r\;,
  \label{eq:rho_lambda}
\end{equation}
where $\ln_+$ denotes the positive branch of the logarithm and where we introduced the same model parameters as for the original K62 model in Eq. (\ref{eq:var_k62}) (see also~\cite{Friedrich_2021} for further information).
Moreover, we introduce Kummer's confluent hypergeometric function
\begin{equation}
  {_1}F_1 (a,b,z)= \frac{\Gamma(b)}{\Gamma(b-a)\Gamma(a)} \int_0^1 \mathrm{d}t e^{zt}t^{a-1}(1-t)^{b-a-1}\;,
\end{equation}
as $\mathcal{K}^H\left(r \right)={_1}F_1 (2H+1,2H+2,r)$, where $\Gamma(z)=\int_0^{\infty} \mathrm{d}t \;t^{z-1}e^{-t}$ denotes the gamma function.

We will now prove that the particular choice (\ref{eq:corr_tensor}) in combination with (\ref{eq:corr_lambda})
ensures that the velocity increment statistics of our model coincides with the K62 prediction. To this end,
we approximate $S_{2\xi,r}(r)\approx \sigma^2 \rho_\xi(r)^{2H}$ in Eq.(\ref{eq:longi_struc}), which is valid for $r \ll L$, and
obtain
\begin{equation}
  S_{n,r}(r)=C_n
  \left(\frac{r+x^\prec}{x^\succ}\right)^{\frac{\mu}{2}(nH- n^2H^2)}r^{nH}\;.
  \label{eq:longi}
\end{equation}
The particular choice $H=1/3$ for the Hurst exponent leads to the well-known K62 model, $ S_{n,r}(r)
\sim r^{\frac{n}{3}-\frac{\mu}{18}n(n-3)}$,
(the same result is obtained by taking the moments of Eq. (\ref{eq:k62}), see also \cite{beck2007statistics} for further proof). Hence, imposing Eq. (\ref{eq:corr_lambda}) for the longitudinal statistics ensures that the random field $\mathbf{u}(\mathbf{x})$ is consistent with the K62 model of turbulence and we have fully specified its joint multipoint statistics (\ref{eq:n_point_3d}) in terms of the model parameters.

\emph{iv.) relations between longitudinal and transverse structure functions:}
As described in Appendix~\ref{app:longi}, we obtain the following expression for the transverse structure functions
\begin{equation}
  S_{n,t}(r)=   \int_0^\infty \mathrm{d}\xi g(\xi) (n-1)!! 2^{\frac{n}{2}} \left[ S_{2\xi,t}(r) \right]^{\frac{n}{2}}\;,
\end{equation}
which can be related to the longitudinal statistics by the von K\'arm\'an-Howarth relation (\ref{eq:karman_naked}) according to
\begin{equation}
  S_{n,t}(r)=  \int_0^\infty \mathrm{d}\xi g(\xi) (n-1)!! 2^{\frac{n}{2}} \left[\frac{1}{2r}\frac{\partial}{\partial r}\left(r^2 S_{2\xi,r}(r) \right) \right]^{\frac{n}{2}}.
\end{equation}
In the inertial range, i.e., for $r \ll L$, this reduces to
\begin{align}
    \lefteqn{S_{n,t}(r)=  \int_0^\infty \mathrm{d}\xi g(\xi) (n-1)!! \sigma^n \rho_\xi(r)^{nH}r^{\frac{n}{2}}}\\ \nonumber
    &~\left[\frac{2H+2}{r}+\frac{\mu H}{r+x^{\prec}}-\frac{\mu H}{r+x^{\prec}} \ln \xi \frac{1}
    {\sqrt{A+\mu \ln_+ \frac{x^{\succ}}{r+x^{\prec}}}} \right]^{\frac{n}{2}}\;.
\end{align}
Here, we can observe that for $\mu =0$, we obtain $S_{n,t}(r) \sim r^{nH}$, i.e., the same power law as for the longitudinal structure functions (\ref{eq:longi}). Whether or not longitudinal and transverse structure functions exhibit different scaling in the inertial range is still not entirely clear~\cite{Shen_2002}. In this model, differences can arise due to the non-Gaussianity
for non-vanishing $\mu$. Further analysis of potential differences between longitudinal and transverse statistics will be covered in a following publication.

Let us briefly summarize the steps that lead to the results obtained in this section: Firstly, we made an ansatz (\ref{eq:corr_tensor}) for the two-point correlation tensor $C_{\xi,\alpha \beta}({\bf r})$, which characterizes the individual Gaussian multivariate statistics in Eq. (\ref{eq:n_point_3d}) and is based on the assumptions of homogeneity and isotropy of the underlying flow. The two remaining scalar functions of the norm of the separation vector $r=|{\bf r}|$, i.e., the longitudinal and transverse correlation functions $C_{\xi,rr}(r)$ and $C_{\xi,tt}(r)$, could further be related by the incompressibility condition for the velocity field, which resulted in Eq. (\ref{eq:karman_naked}). Finally, the longitudinal statistics (\ref{eq:longi}) was fixed by the K62-scaling, which was achieved by imposing Eq. (\ref{eq:corr_lambda}) for the longitudinal correlations  $C_{\xi,rr}(r)$.
In the following section, we discuss a sampling algorithm that allows one to draw samples from the joint multipoint PDF (\ref{eq:n_point_3d}).

\subsection{Sampling algorithm for superstatistical random fields}
\label{sec:sample}
In general, sampling from the joint $n$-point PDF (\ref{eq:n_point_3d}) is a rather challenging task and one typically has to resort to Markov Chain Monte Carlo methods or collapsed Gibbs sampling. In the following, we propose a sampling algorithm which is motivated by sampling of an ordinary Gaussian random field in Fourier space. The central observation here is that
Eq. (\ref{eq:n_point_3d}) can be interpreted as a \emph{scale mixture} of Gaussians with different covariances. Therefore, we can first construct a $3+1$-dimensional Gaussian velocity field $u_\alpha(\xi, \mathbf{x})$ where the parameter $\xi$ can be interpreted as an additional time coordinate.

The scale mixture can then be introduced by assigning each point in space $\mathbf{x}$ to a different parameter $\xi$, weighed by the lognormal distribution.
In some sense, the procedure is thus reminiscent of the sampling of a subordinated process in the context of the Continuous Time Random Walk (indeed, it can be shown that the re-parameterized scale $\rho_\xi(r)$ in Eq. (\ref{eq:rho_lambda}) obeys a Langevin equation with multiplicative noise). On the basis of these observation our sampling algorithm can roughly be divided into two steps:

\emph{i.)} A Gaussian random field $u_\alpha(\xi, \mathbf{x})$ with an additional dependence on the parameter $\xi$ is constructed in the usual way in Fourier space
$\hat u_\alpha(\xi, \mathbf{k})= \int \textrm{d}\mathbf{x} e^{-i \mathbf{k}\cdot \mathbf{x}}u_\alpha(\xi, \mathbf{x})$. For a homogeneous and isotropic velocity field, the
kinetic energy spectrum $E_\xi(k)$ can be obtained from a Fourier transform of Eq. (\ref{eq:corr_tensor}), i.e., $\hat   C_{\xi,\alpha \beta}(\mathbf{k})= \int \textrm{d}\mathbf{r} e^{-i \mathbf{k}\cdot \mathbf{r}}  C_{\xi,\alpha \beta}(\mathbf{r})$ according to
$E_\xi(k)= \frac{1}{2\pi k^2} \sum_{\alpha=\beta}C_{\xi,\alpha \beta}(\mathbf{k})$. A Gaussian random field with the desired kinetic energy spectrum $E(\xi,k)$ can now be constructed as
\begin{equation}
 \hat u_\alpha(\xi, \mathbf{k})= E_\xi^{\frac{1}{2}}(k) \varphi_{\alpha}(\mathbf{ k})\;,
 \label{eq:gauss_rand}
\end{equation}
where $\boldsymbol{\varphi}(\mathbf{k})$ is a white noise vector with random phases and unit amplitude, which satisfies
$\langle  \varphi_{\alpha}(\mathbf{ k}) \varphi_{\beta}(\mathbf{ k}')\rangle=\frac{1}{4\pi k^2}\left(-\frac{k_\alpha k_\beta}{k^2}+\delta_{\alpha \beta} \right)\delta(\mathbf{k}+\mathbf{k}')$.
Here, it is important to stress that the same realization $\varphi_{\alpha}(\mathbf{ k})$ has to be chosen for each parameter $\xi$ in order to guarantee the appropriate scale mixture.

\emph{ii.)} After assembling $\hat u_\alpha(\xi, \mathbf{k})$ according to Eq. (\ref{eq:gauss_rand}) and an inverse Fourier transform, the desired random field $u_\alpha(\mathbf{x})$ is obtained by assigning each spatial point $\mathbf{x}$ a different value $\xi$ from the family of Gaussian random fields $u_\alpha(\xi, \mathbf{x})$. Here, one has to guarantee that neighboring points of ${\bf x}$ have been assigned similar values of $\xi$ and that the change in $\xi$ is not too abrupt (this ensures that one does not
converge against fractional Gaussian noise at small scale separations).
The sampling algorithm is also summarized in Appendix~\ref{app:sample}.

Hence, the proposed sampling method relies on the assumed equivalence of the ensemble average in Eq. (\ref{eq:n_point_3d}) (i.e., the average over Gaussian ensembles characterized by different covariance matrices $\sigma_{\xi}$) and
the average over the reference point $\mathbf{x}$, which is usually invoked to calculate statistical quantities (e.g., correlation or structure functions) from the random field $\mathbf{u}(\mathbf{x})$.

The advantage of this method is that one avoids the full computation of the covariance matrix in Eq. (\ref{eq:n_point_3d}) which is typically required for Monte Carlo-type sampling methods. Therefore, the computational costs
related to the random field synthesis by this method are comparable to the spectral models by Mann or Veers.
In the following section, we further outline a method which integrates point-wise turbulence measurement data sets into such superstatistical random fields.
\begin{figure*}[ht]
\centering
\includegraphics[width=0.48 \textwidth]{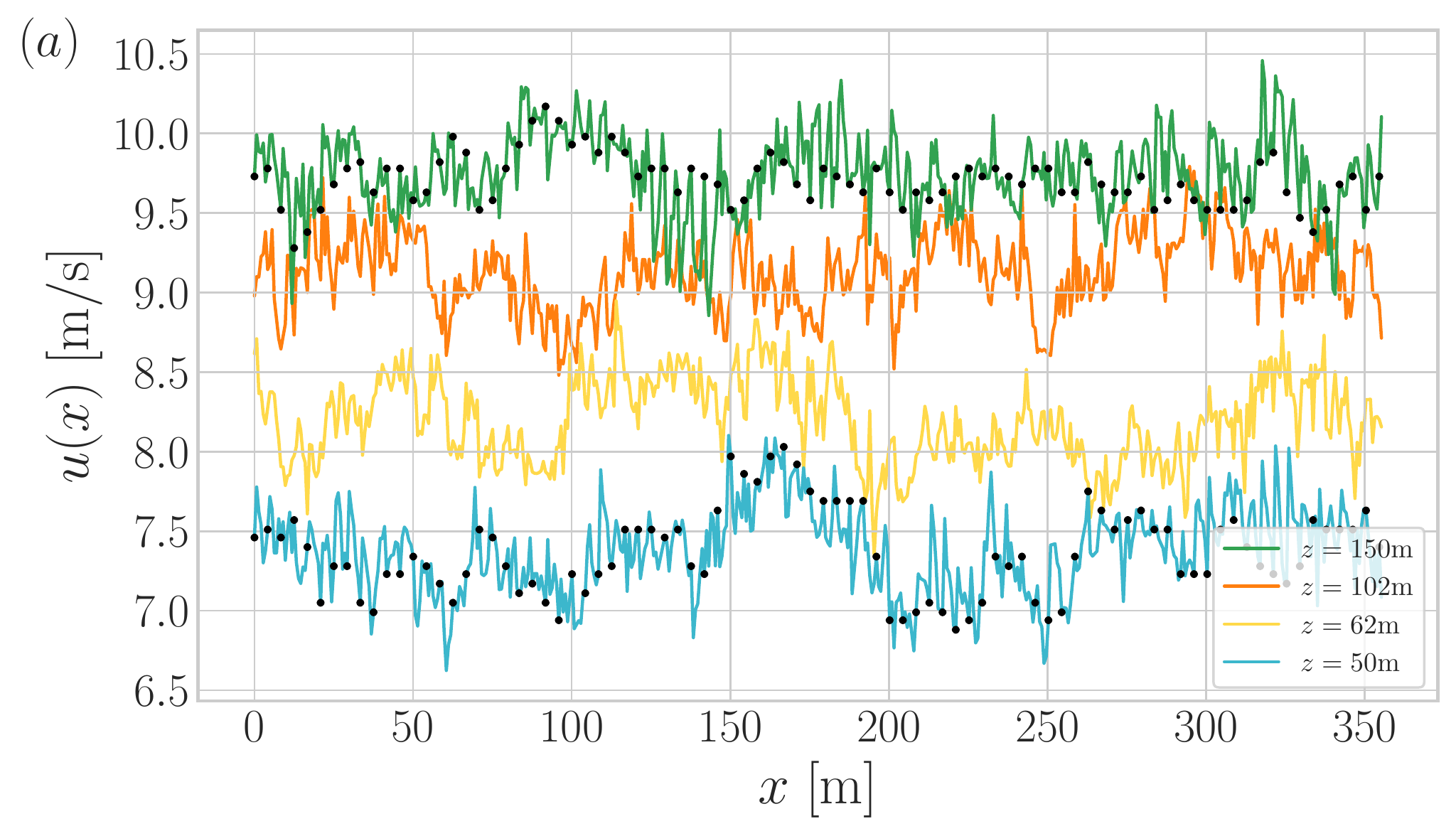}
\hspace{0.1cm}
\includegraphics[width=0.47 \textwidth]{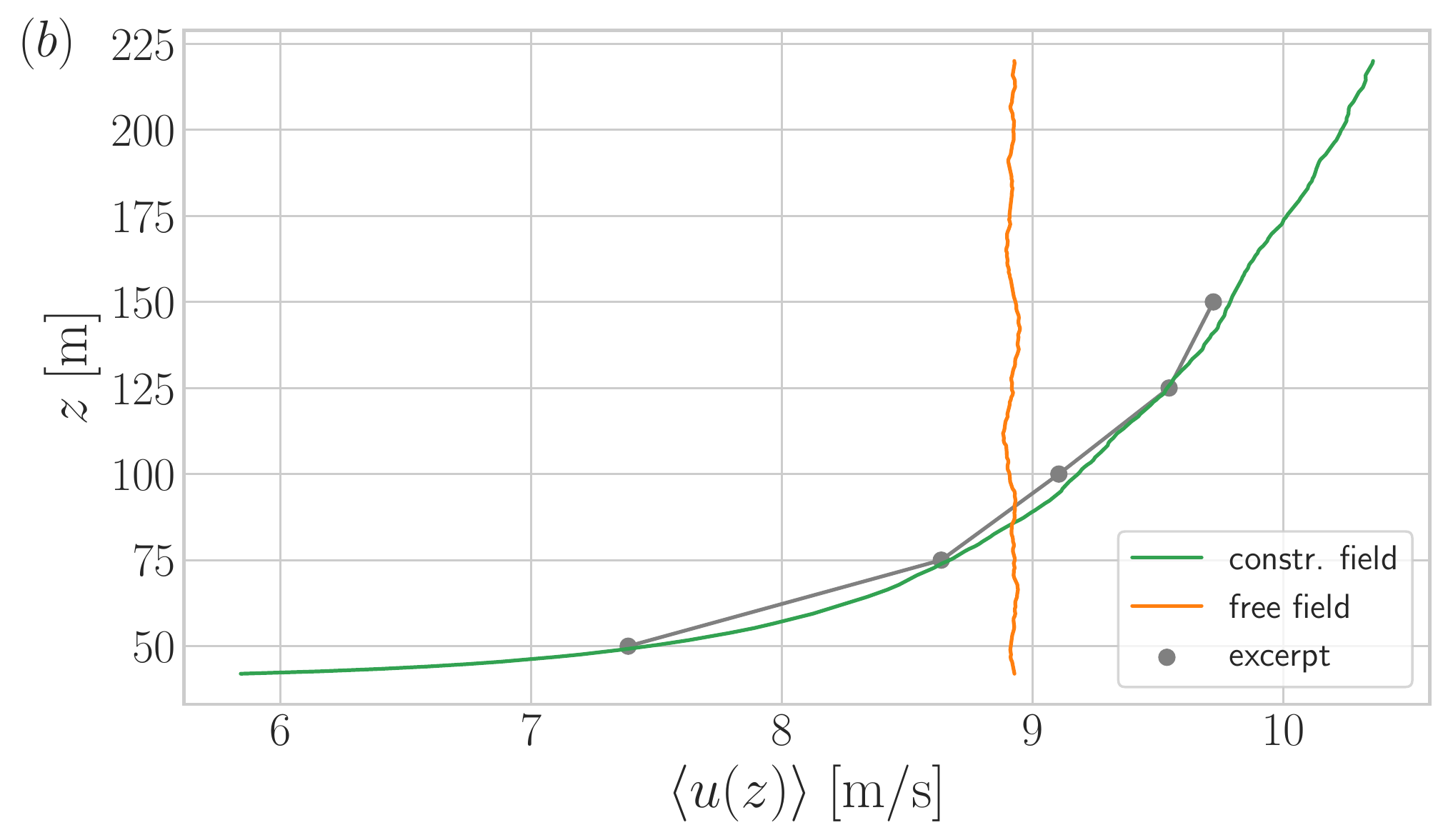}
\caption{(a) Extracts from the superstatistical random field for fixed $y=-14\textrm{m}$ and at three different heights $z=[50, 100, 150] \textrm{m}
$. The black points indicate the original GROWIAN time series which are exactly hit by the stochastic interpolation. The orange curve corresponds to the hub height which is not covered by the measurement array. (b) Vertical velocity profile of the superstatistical random field (blue) from averaging the entire field in $x-$ and $y-$direction.}
\label{fig:interpolated_data}
\end{figure*}

\subsection{Stochastic interpolation of point-wise atmospheric turbulence data by superstatistical random fields}
\label{sec:interpol}
The superstatistical random fields of the previous section can be considered as homogeneous isotropic turbulent wind fields with parameters that can be determined empirically (intermittency coefficient $\mu$, Hurst exponent $H$, and integral length scale $L$). In this section, we discuss a modification of the sampling algorithm that is capable of including experimental measurements $\mathbf{U}_i$ at points $\mathbf{x}_i$.
In order to constrain the superstatistical random field $\mathbf{u}(\mathbf{x})$ on these points, we apply the methodology of multipoint Gaussian bridge processes~\cite{friedrich2020stochastic,bierbooms2009constrained} for the family of Gaussian random fields $\mathbf{u}(\xi, \mathbf{x})$ in the previous section.  Here, the point-wise measurements are denoted by $\mathbf{U}_i$
at points $\mathbf{x}_i$. The bridge process can thus be constructed from $\mathbf{u}(\xi, \mathbf{x})$
according to
\begin{align}
  \lefteqn{\mathbf{u}^B(\xi,\mathbf{x})}
  \\ \nonumber
   =& \mathbf{u}(\xi,\mathbf{x})- \left[u_\alpha(\xi,\mathbf{x}_i)-U_{\alpha,i} \right] \sigma_{\xi,i\alpha;j\beta}^{-1} \left \langle u_\beta(\xi,\mathbf{x}_j) \mathbf{u}(\xi,\mathbf{x}) \right \rangle\;,
   \label{eq:bridge}
\end{align}
where the covariance matrix is defined as $\sigma_{\xi,i\alpha;j\beta}=\left \langle
u_\alpha(\mathbf{x}_i) u_\beta(\mathbf{x}_j) \right \rangle$ and summation over the same indices is implied. It can be seen that each bridge process (for varying $\xi$) exactly ``hits'' the prescribed measurement points, i.e.,
\begin{align}
  \lefteqn{{u}_\gamma^B(\xi,\mathbf{x}_k)}\\ \nonumber
   =& u_\gamma(\xi,\mathbf{x}_k)-
  \left[u_\alpha(\xi,\mathbf{x}_i)-U_{\alpha,i} \right] \sigma_{\xi,i\alpha;j\beta}^{-1} \underbrace{\left \langle u_\beta(\xi,\mathbf{x}_j)
  {u}_\gamma(\xi,\mathbf{x}_k) \right \rangle}_{\sigma_{\xi,j \beta;k \gamma}} \\ \nonumber
  =& u_\gamma(\xi,\mathbf{x}_k)-
 \left[u_\alpha(\xi,\mathbf{x}_i)-U_{\alpha,i} \right] \delta_{ik}\delta_{\alpha\gamma}= U_{\gamma,k}\;, 
\end{align}
where $\delta_{ij}$ denotes the Kronecker delta.

Non-Gaussian features are again generated by subsequent scale mixing at each point $\mathbf{x}$ with different $\xi$, as outlined in the previous section (see also Appendix~\ref{app:sample}). 
Hence, combining the Gaussian bridge processes (\ref{eq:bridge}) with the proposed non-Gaussian sampling algorithm in Sec.~\ref{sec:sample} provides a three-dimensional reconstruction of a velocity field with exactly controllable intermittency properties from a sparse measurement set. In the following section, we will give a  more detailed account of this integrated reconstruction method at the example of the GROWIAN measurement and the extracted model parameters from Sec.~\ref{sec:charac}.

\section{Results and validation for the reconstructed superstatistical wind fields}
In this section, we demonstrate our reconstruction method at the example of the GROWIAN measurements. Furthermore, we perform a model validation by comparing certain statistical quantities (e.g., correlation functions, flatness, or longitudinal velocity increment PDFs) against empirical quantities. We must emphasize that this reconstruction method assumes the universality of small-scale turbulent fluctuations in the sense of a K62-type model (i.e., Fig.~\ref{fig:growian_data}(d) should be
continuable to arbitrarily small scales). Nonetheless, the model is finely tunable for any inflow condition and site, for which model parameters have to be determined by the procedure outlined in Sec.~\ref{sec:charac}.
\begin{table}[h!]  
  \begin{center}
    \begin{tabular}{l l l l l}
 	dx [m] &  $H$ &  $\mu$ & $L$ [km] & $u_{rms}$ [m/s]\\
	\hline 0.654 $\qquad$ & 0.338 $\qquad$ & 0.243 $\qquad$ & 10.75 $\qquad$ & 1.048
    \end{tabular}
  \end{center}
  \caption{Recapitulation of model parameters determined from the GROWIAN data set (see Sec. \ref{sec:charac} for further details): Spatial resolution dx, Hurst exponent $H$, intermittency coefficient $\mu$, integral length scale $L$, and the standard deviation of the time series extract in Fig.~\ref{fig:growian_data}(a) which fixes the remaining model parameter $\sigma$ through Eq. (\ref{eq:corr_lambda}) according to $u_{rms}=\sigma L^H \sqrt{\Gamma(2H+1)/2}$.}
\end{table}

To this end, we generated a full three-dimensional velocity field from an exemplary data set of the GROWIAN measurements (see gray-shaded area in Fig.~\ref{fig:growian_data}(a)) with a resolution of $256^2\times 768$ grid points. Here, we reconstructed the velocity field component in the direction of mean wind speed ($x$-component), as the propeller anemometers do not measure all three components simultaneously. Moreover, our method fills up the velocity field with zero-mean fluctuations and we have to first subtract the mean vertical velocity profile from the measurement points.
As we also fill up points in between the five heights in Fig.~\ref{fig:growian_data}(b), we apply a fit for the shear profile (we assumed a logarithmic profile) and then subtract this profile from the measurement points $U_{\alpha,i}$
at the corresponding heights. In the last step the mean vertical velocity profile is added to the stochastically interpolated velocity field. Further model parameters ($H,\mu,L,\sigma$) have to be determined from measurement data (see also Sec.~\ref{sec:charac}) and are further summarized in Tab.~I. Here, the Hurst exponent $H$, the intermittency coefficient $\mu$, and the integral length scale $L$ have been determined from a global analysis of the entire data set (by fitting the PDFs in Fig.~\ref{fig:growian_data}(d)), whereas the parameter $\sigma$ entering the correlation function (\ref{eq:corr_lambda}) has been obtained from the standard deviation $u_{rms}$ of the time series extract (gray-shaded area in Fig.~\ref{fig:growian_data}(a)) by  $u_{rms}=\sigma L^H \sqrt{\Gamma(2H+1)/2}$ (please note that the dimension of $\sigma$ is [m/$\textrm{s}^2$] only for the special case of $H=0.5$). Furthermore, the model parameter $L$ does not constitute the integral length scale of the data set, but is a fit parameter which is consistent with the small-scale evolution of the velocity increment PDFs in Fig.~\ref{fig:growian_data}(d).

\begin{figure}[ht]
\centering
\includegraphics[width=0.48 \textwidth]{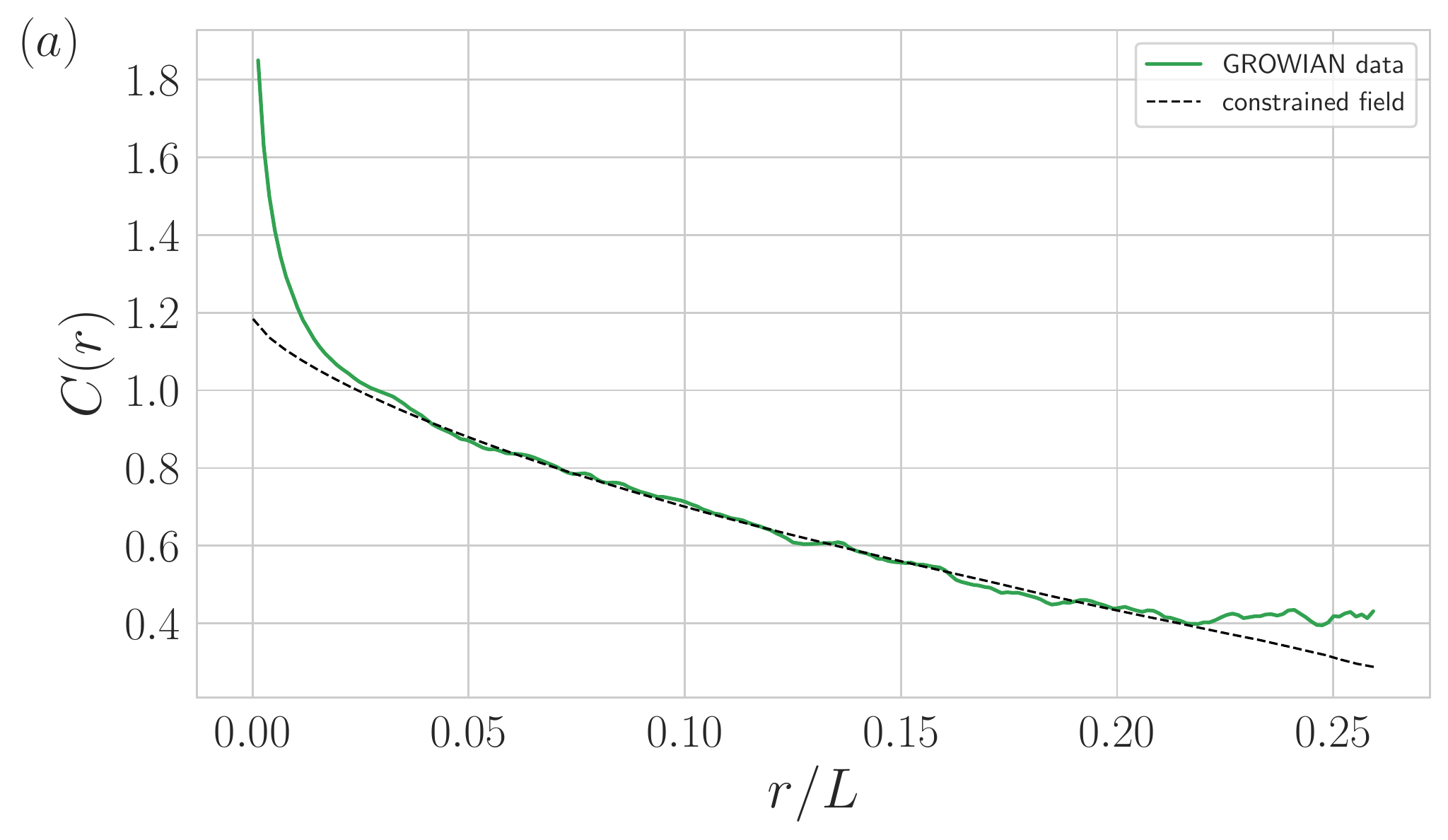}
\includegraphics[width=0.485 \textwidth]{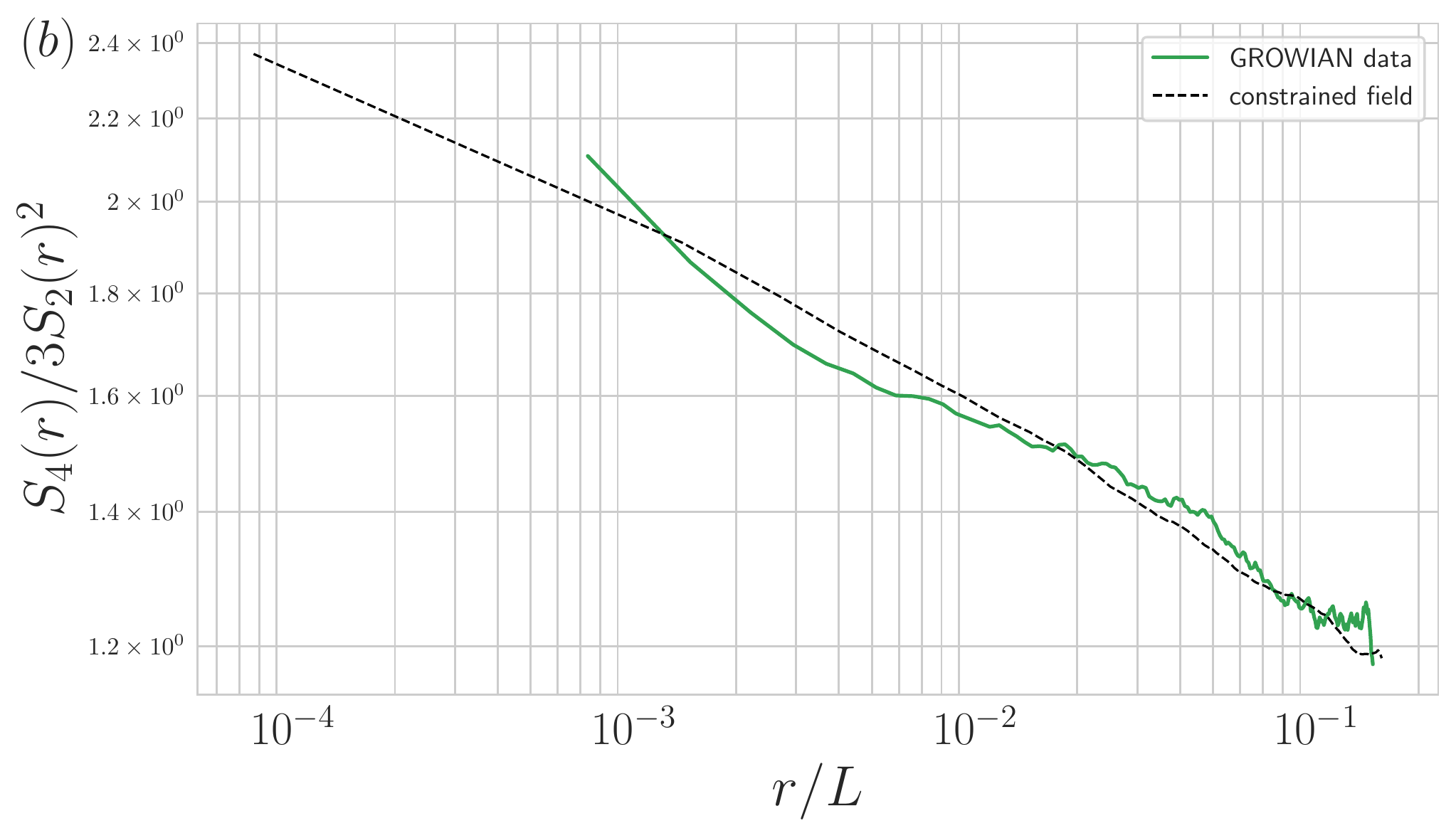}
\includegraphics[width=0.485 \textwidth]{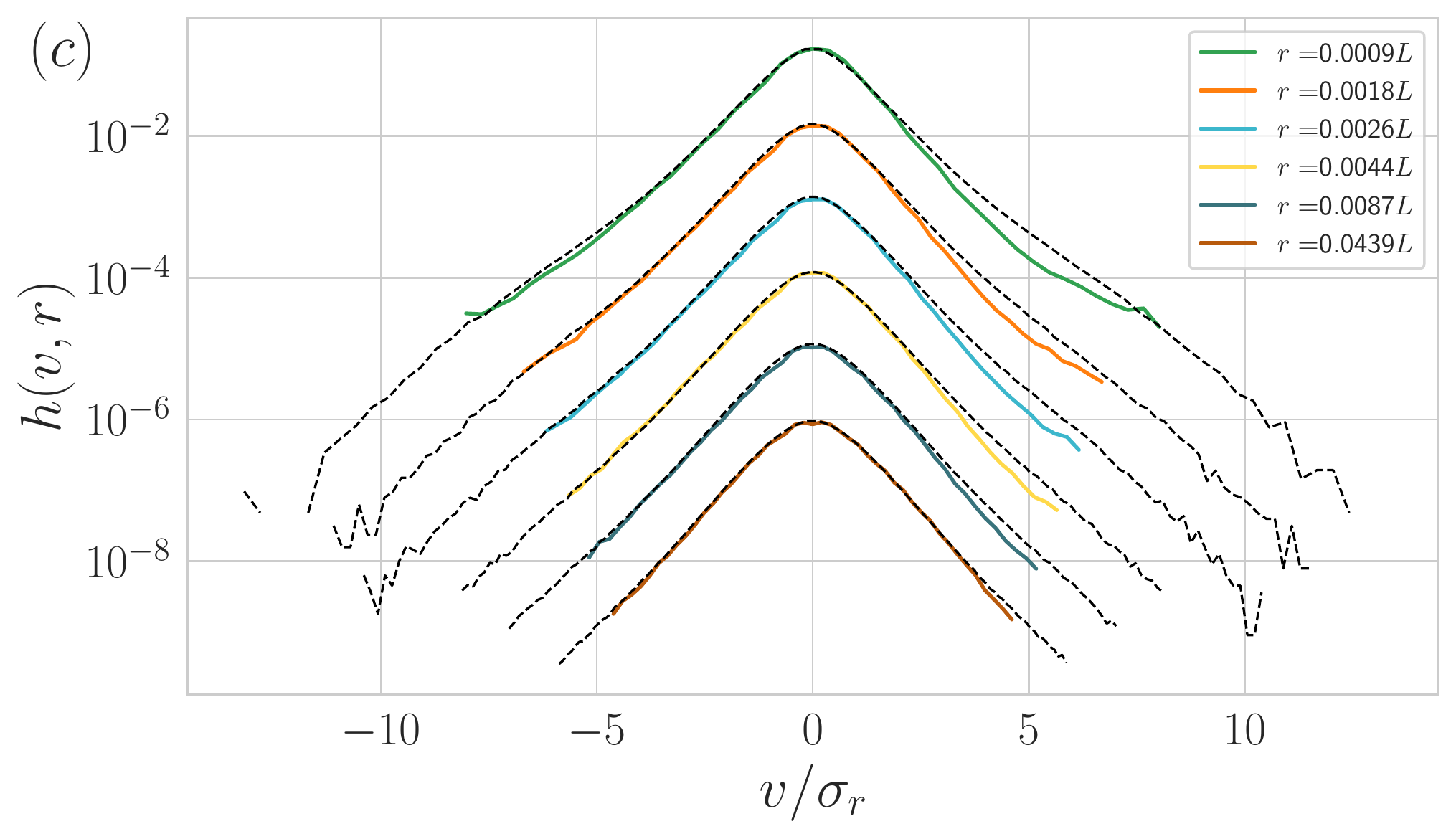}
\caption{(a) Correlation functions $C(r)$ evaluated from the entire GROWIAN data set (green). The dashed lines corresponds to the correlation function of the reconstructed superstatistical random field depicted in Fig.~\ref{fig:scatter}. (b) Flatness $S_4(r)/3 S_2(r)^2$ of the GROWIAN data set (green) as well as of the superstatistical random field (dashed curve). A flatness of $1$ would be tantamount to a Gaussian distribution of velocity increments. (c) Velocity increment PDFs of the GROWIAN data set (same PDFs as shown in Fig.~\ref{fig:growian_data}(d)). Dashed lines correspond to the reconstructed superstatistical random field.}
\label{fig:interpolated_data_stat}
\end{figure}
A scatter plot for a certain range of the reconstructed superstatistical velocity field is shown in Fig.~\ref{fig:scatter} and exhibits a rather convoluted structure. Interestingly, some puffs of comparably low velocity can be found at larger heights, but it is not evident whether these result
from the constraining on the GROWIAN data or if they are an intrinsic feature of the superstatistical random field itself.
A slice through the velocity field $u(\mathbf{x})$ for fixed $y=-14$ m and four different heights $z$ is depicted in Fig.~\ref{fig:interpolated_data}(a).
The black dots indicate the original measurement points and are exactly hit by the reconstructed field, which is a direct consequence of the bridge construction in Eq. (\ref{eq:bridge}).
Fig.~\ref{fig:interpolated_data_stat}(b) depicts the vertical velocity profile of the velocity field, where the gray points
correspond to the measurements (black points in  Fig.~\ref{fig:interpolated_data}(a)). 
Here, the vertical velocity profile of the reconstructed field (green) is in good
agreement with the measurements (gray curve). For comparison, we also show the vertical velocity profile of the ``free velocity field'' which has been obtained from the superstatistical random field that enters the r.h.s. of the bridge construction in Eq. (\ref{eq:bridge}). As expected, the free field is un-sheared, therefore, shear in the green curve is a direct consequence of the measurement data entering the bridge construction.

Fig.~\ref{fig:interpolated_data_stat}(a) depicts the correlation function $C(r)$ of the  entire GROWIAN data set (green) and agrees fairly well with the correlation function of the reconstructed superstatistical random field (dashed curve). The deviations for small $r$ are due to different standard deviations of the velocities evaluated from the entire data set and the excerpt (gray-shaded area in Fig.~\ref{fig:growian_data}(a)). Moreover, we emphasize that - although the underlying wind field model implies the assumption of homogeneity -  we recover the decay of correlations at scales $r < 0.25 L$ quite well.
Fig.~\ref{fig:interpolated_data_stat}(b) shows the flatness $S_4(r)/3S_2(r)^2$ which characterizes deviations from Gaussian distributions of the velocity increments (Gaussianity would imply a flatness of $1$). The flatness increases with decreasing $r$ and the GROWIAN predictions (green curve) and the superstatistical random field agree well. Both curves deviate only for small $r$ where the GROWIAN data set exhibits deviations from the nearly linear behavior of the reconstructed field (Fig.~\ref{fig:interpolated_data_stat}(b) is plotted in log-log-representation).
The latter behavior might be attributed to the limited temporal resolution of the propeller anemometers or might indicate the presence of a dissipation range, and is thus difficult to assess any further. 
The statistics of the longitudinal velocity increments of the reconstructed superstatistical random field  is depicted  Fig.~\ref{fig:interpolated_data}(c)
in form of the dashed lines. Due to the exactly controllable intermittency properties of the random field (by virtue of the joint multipoint statistics (\ref{eq:fine-n-point}) in combination with the model calibration with the parameters in Tab.~I, the superstatistical random field reproduces the non-Gaussian properties of the GROWIAN data set (colored curves). This can also be proven by calculating the modified $n$-point statistics (\ref{eq:n_point_3d}) for the reconstructed velocity field $\mathbf{u}^B(\mathbf{x})$ as shown in Appendix~\ref{app:modified}.

%

\section{Conclusions and potential model improvements}
\label{sec:improve}
We have presented a novel method to reconstruct a non-Gaussian velocity field from a set of sparse, point-wise atmospheric turbulence measurements. The method is highly relevant for many applications such as the estimation of loads on wind turbines~\cite{muecke}, to refine meso-scale models of atmospheric turbulence, as well as for reconstructions of temperature fields or aerosol concentrations. In contrast, to commonly used wind field models~\cite{veers1984modeling,MANN1998269}, our method controls the empirically observed intermittency of atmospheric turbulence with very high accuracy. To our knowledge, the combination of non-Gaussian random fields and the stochastic interpolation of a certain number of sparse measurement points has never been proposed before and should lead to new research collaborations between experiment, numerical simulations, and theory.
As stated in Sec.~\ref{sec:superstat} on superstatistical random field synthesis, our model is currently set up in terms of a homogeneous isotropic turbulence and thus
neglects small-scale statistical features due to shear. Only the subsequent reconstruction on the basis of the meteorological mast measurements incorporates the effect of shear, small-scale fluctuations, however, remain homogeneous and isotropic. We must stress that this can be considered as a zeroth-order approximation, and future work has to be devoted to the important question whether shear also acts on small-scales. It seems plausible to assume that the effect is scale-dependent and one observes a critical balance between
horizontal and vertical fluctuations~\cite{nazarenko_schekochihin_2011}. If the latter holds true, one should directly modify the correlation tensor (\ref{eq:corr_tensor}). Due to the fact that the proposed statistics in Sec.~\ref{sec:superstat} consists of a superposition of Gaussian statistics,
it is possible to deal with shear and atmospheric stability by similar concepts as in the case of the Mann model~\cite{chougule2017modeling,chougule2018simplification}.
Another possibility
would be to use the tensorial form of axisymmetric turbulence~\cite{1950,Robertson1940} with respect to a preferred direction $\boldsymbol{\lambda}$, namely
\begin{align}\nonumber
  \lefteqn{C_{\xi,\alpha\beta}(\mathbf{r},\boldsymbol{\lambda})= A_\xi(r,\lambda) \frac{r_\alpha r_\beta}{r^2}+B_\xi(r,\lambda) \delta_{\alpha \beta}}\\
   &+ C_\xi(r,\lambda) \frac{r_\alpha \lambda_\beta}{r \lambda}+D_\xi(r,\lambda) \frac{r_\alpha r_\beta}{r^2}+E_\xi(r,\lambda) \frac{\lambda_\alpha \lambda_\beta}{\lambda^2}\;.
   \label{eq:axi_tensor}
\end{align}
Latter tensorial form involves five unknown scalar functions, which can be reduced to four by imposing the incompressibility condition of the velocity field. Therefore, other hypothesis, e.g., a scale-dependent balance between the cross terms in Eq. (\ref{eq:axi_tensor}), have to be invoked. Another assumption that has been invoked for the reconstruction of the superstatical random field was Taylor's hypothesis. Numerous
studies have shown that the use of Taylor's hypothesis in the atmospheric surface layer can be problematic~\cite{cheng2017failure,del2009estimation,yang2018implication}. Therefore, future work
has to be devoted to a more comprehensive assessment of Taylor's hypothesis in the context of the wind field reconstruction. However, it should be possible to account for a spatio-temporal spectrum of turbulence by the use of Kraichnan's random sweeping hypothesis~\cite{wilczek2015spatio} where the correlation tensor (\ref{eq:corr_tensor}) should be modified accordingly.

From a numerical point of view, the proposed algorithm in Sec.~\ref{sec:sample} for the construction of purely superstatistical random fields, i.e., without constraining on data points, is rather effective as it operates directly in Fourier space. The resolution of the random fields is therefore limited by machine memory. Only the subsequent reconstruction (\ref{eq:bridge}) imposes difficulties as it involves the calculation of correlations between all grid points with the prescribed points (last correlator in Eq. (\ref{eq:bridge})). Latter issue, however, can be addressed by tensor decomposition such as matrix product states. Increasing the wind field reconstruction interval in Fig.~\ref{fig:growian_data}(a) will be a necessary prerequisite for further applications of superstatistical random fields. This applies in particular to the accurate load or power curve estimations in the context of the wind energy sciences for which statistical significance is a key issue. It is also imaginable that the proposed scale refinement, which ``injects'' arbitrarily small turbulent fluctuations into a large-scale flow configuration, could be used as a stochastic subgrid model in the context of large eddy simulations. Nonetheless, latter approach will require further considerable efforts in making the constraining algorithm and the Fourier-based scale mixing more computationally efficient.

Further potential future applications include the small-scale enhancement of LiDAR measurements, where small-scale turbulent fluctuations are averaged over probe volumes~\cite{amt-15-1355-2022,mikkelsen2014lidar}, as well as the study of particle transport in the here-proposed synthetic fields. Furthermore, motivated by studies of wind speed persistence in the atmosphere~\cite{weber2019wind,Chowdhuri_2020}, it would be interesting to investigate whether the here-proposed wind fields exhibit similar features and how these might be connected to wind shear. In this context, we stress again that the current work was motivated by the modeling of small-scale wind field fluctuations by a homogeneous wind field model and future work has to be devoted to account for large-scale coherent structures and memory effects in the atmospheric boundary layer as well~\cite{Laudani_2021}.
 As far as basic turbulence research is concerned, the proposed joint multipoint statistics could also be applied to the hierarchical problem in the statistical description of the Navier-Stokes equation~\cite{friedrich2020non}. Latter approach could yield important insights in the physical mechanism underlying the phenomenon of small-scale intermittency, which is not captured by conventional Gaussian approximations.
\section{Acknowledgements}
J. F. is grateful for fruitful discussion with J. Mann.
This work is partly funded by the German Federal Ministry for Economic Affairs and Energy
in the scope of the projects EMUwind (03EE2031A/C) and PASTA (03EE2024A/B).
J. F. acknowledges funding from the Humboldt Foundation within a Feodor-Lynen fellowship.
\bibliography{wind_fields.bib}

\begin{thebibliography}{76}%
\makeatletter
\providecommand \@ifxundefined [1]{%
 \@ifx{#1\undefined}
}%
\providecommand \@ifnum [1]{%
 \ifnum #1\expandafter \@firstoftwo
 \else \expandafter \@secondoftwo
 \fi
}%
\providecommand \@ifx [1]{%
 \ifx #1\expandafter \@firstoftwo
 \else \expandafter \@secondoftwo
 \fi
}%
\providecommand \natexlab [1]{#1}%
\providecommand \enquote  [1]{``#1''}%
\providecommand \bibnamefont  [1]{#1}%
\providecommand \bibfnamefont [1]{#1}%
\providecommand \citenamefont [1]{#1}%
\providecommand \href@noop [0]{\@secondoftwo}%
\providecommand \href [0]{\begingroup \@sanitize@url \@href}%
\providecommand \@href[1]{\@@startlink{#1}\@@href}%
\providecommand \@@href[1]{\endgroup#1\@@endlink}%
\providecommand \@sanitize@url [0]{\catcode `\\12\catcode `\$12\catcode
  `\&12\catcode `\#12\catcode `\^12\catcode `\_12\catcode `\%12\relax}%
\providecommand \@@startlink[1]{}%
\providecommand \@@endlink[0]{}%
\providecommand \url  [0]{\begingroup\@sanitize@url \@url }%
\providecommand \@url [1]{\endgroup\@href {#1}{\urlprefix }}%
\providecommand \urlprefix  [0]{URL }%
\providecommand \Eprint [0]{\href }%
\providecommand \doibase [0]{https://doi.org/}%
\providecommand \selectlanguage [0]{\@gobble}%
\providecommand \bibinfo  [0]{\@secondoftwo}%
\providecommand \bibfield  [0]{\@secondoftwo}%
\providecommand \translation [1]{[#1]}%
\providecommand \BibitemOpen [0]{}%
\providecommand \bibitemStop [0]{}%
\providecommand \bibitemNoStop [0]{.\EOS\space}%
\providecommand \EOS [0]{\spacefactor3000\relax}%
\providecommand \BibitemShut  [1]{\csname bibitem#1\endcsname}%
\let\auto@bib@innerbib\@empty
\bibitem [{\citenamefont {Wyngaard}(1992)}]{wyngaard1992atmospheric}%
  \BibitemOpen
  \bibfield  {author} {\bibinfo {author} {\bibfnamefont {J.~C.}\ \bibnamefont
  {Wyngaard}},\ }\bibfield  {title} {\bibinfo {title} {Atmospheric
  turbulence},\ }\href@noop {} {\bibfield  {journal} {\bibinfo  {journal}
  {Annual Review of Fluid Mechanics}\ }\textbf {\bibinfo {volume} {24}},\
  \bibinfo {pages} {205} (\bibinfo {year} {1992})}\BibitemShut {NoStop}%
\bibitem [{\citenamefont {Sutton}(2020)}]{sutton2020atmospheric}%
  \BibitemOpen
  \bibfield  {author} {\bibinfo {author} {\bibfnamefont {O.~G.}\ \bibnamefont
  {Sutton}},\ }\href@noop {} {\emph {\bibinfo {title} {Atmospheric
  turbulence}}}\ (\bibinfo  {publisher} {Routledge},\ \bibinfo {year}
  {2020})\BibitemShut {NoStop}%
\bibitem [{\citenamefont {Richardson}(1921)}]{richardson1921some}%
  \BibitemOpen
  \bibfield  {author} {\bibinfo {author} {\bibfnamefont {L.~F.}\ \bibnamefont
  {Richardson}},\ }\bibfield  {title} {\bibinfo {title} {I. {S}ome measurements
  of atmospheric turbulence},\ }\href@noop {} {\bibfield  {journal} {\bibinfo
  {journal} {Phil. Trans. R. Soc. A}\ }\textbf {\bibinfo {volume} {221}},\
  \bibinfo {pages} {1} (\bibinfo {year} {1921})}\BibitemShut {NoStop}%
\bibitem [{\citenamefont {Richardson}(2007)}]{richardson2007weather}%
  \BibitemOpen
  \bibfield  {author} {\bibinfo {author} {\bibfnamefont {L.~F.}\ \bibnamefont
  {Richardson}},\ }\href@noop {} {\emph {\bibinfo {title} {Weather prediction
  by numerical process}}}\ (\bibinfo  {publisher} {Cambridge university
  press},\ \bibinfo {year} {2007})\BibitemShut {NoStop}%
\bibitem [{\citenamefont {Monin}(1958)}]{Monin_1958}%
  \BibitemOpen
  \bibfield  {author} {\bibinfo {author} {\bibfnamefont {A.~S.}\ \bibnamefont
  {Monin}},\ }\bibfield  {title} {\bibinfo {title} {The structure of
  atmospheric turbulence},\ }\href {https://doi.org/10.1137/1103023} {\bibfield
   {journal} {\bibinfo  {journal} {Theory of Probability {\&} Its
  Applications}\ }\textbf {\bibinfo {volume} {3}},\ \bibinfo {pages} {266}
  (\bibinfo {year} {1958})}\BibitemShut {NoStop}%
\bibitem [{\citenamefont {Oboukhov}(1962)}]{Oboukhov1962}%
  \BibitemOpen
  \bibfield  {author} {\bibinfo {author} {\bibfnamefont {A.~M.}\ \bibnamefont
  {Oboukhov}},\ }\bibfield  {title} {\bibinfo {title} {{Some specific features
  of atmospheric turbulence}},\ }\href@noop {} {\bibfield  {journal} {\bibinfo
  {journal} {J. Fluid Mech.}\ }\textbf {\bibinfo {volume} {67}},\ \bibinfo
  {pages} {77} (\bibinfo {year} {1962})}\BibitemShut {NoStop}%
\bibitem [{\citenamefont {Morales}\ \emph {et~al.}(2012)\citenamefont
  {Morales}, \citenamefont {W{\"a}chter},\ and\ \citenamefont
  {Peinke}}]{morales2012characterization}%
  \BibitemOpen
  \bibfield  {author} {\bibinfo {author} {\bibfnamefont {A.}~\bibnamefont
  {Morales}}, \bibinfo {author} {\bibfnamefont {M.}~\bibnamefont
  {W{\"a}chter}},\ and\ \bibinfo {author} {\bibfnamefont {J.}~\bibnamefont
  {Peinke}},\ }\bibfield  {title} {\bibinfo {title} {Characterization of wind
  turbulence by higher-order statistics},\ }\href@noop {} {\bibfield  {journal}
  {\bibinfo  {journal} {Wind Energy}\ }\textbf {\bibinfo {volume} {15}},\
  \bibinfo {pages} {391} (\bibinfo {year} {2012})}\BibitemShut {NoStop}%
\bibitem [{\citenamefont {W{\"a}chter}\ \emph {et~al.}(2012)\citenamefont
  {W{\"a}chter}, \citenamefont {Hei{\ss}elmann}, \citenamefont {H{\"o}lling},
  \citenamefont {Morales}, \citenamefont {Milan}, \citenamefont {M{\"u}cke},
  \citenamefont {Peinke}, \citenamefont {Reinke},\ and\ \citenamefont
  {Rinn}}]{wachter2012turbulent}%
  \BibitemOpen
  \bibfield  {author} {\bibinfo {author} {\bibfnamefont {M.}~\bibnamefont
  {W{\"a}chter}}, \bibinfo {author} {\bibfnamefont {H.}~\bibnamefont
  {Hei{\ss}elmann}}, \bibinfo {author} {\bibfnamefont {M.}~\bibnamefont
  {H{\"o}lling}}, \bibinfo {author} {\bibfnamefont {A.}~\bibnamefont
  {Morales}}, \bibinfo {author} {\bibfnamefont {P.}~\bibnamefont {Milan}},
  \bibinfo {author} {\bibfnamefont {T.}~\bibnamefont {M{\"u}cke}}, \bibinfo
  {author} {\bibfnamefont {J.}~\bibnamefont {Peinke}}, \bibinfo {author}
  {\bibfnamefont {N.}~\bibnamefont {Reinke}},\ and\ \bibinfo {author}
  {\bibfnamefont {P.}~\bibnamefont {Rinn}},\ }\bibfield  {title} {\bibinfo
  {title} {The turbulent nature of the atmospheric boundary layer and its
  impact on the wind energy conversion process},\ }\href@noop {} {\bibfield
  {journal} {\bibinfo  {journal} {J. Turbul.}\ ,\ \bibinfo {pages} {N26}}
  (\bibinfo {year} {2012})}\BibitemShut {NoStop}%
\bibitem [{\citenamefont {Mikkelsen}(2014)}]{mikkelsen2014lidar}%
  \BibitemOpen
  \bibfield  {author} {\bibinfo {author} {\bibfnamefont {T.}~\bibnamefont
  {Mikkelsen}},\ }\bibfield  {title} {\bibinfo {title} {Lidar-based research
  and innovation at dtu wind energy--a review},\ }\href@noop {} {\bibfield
  {journal} {\bibinfo  {journal} {J. Phys. Conf. Ser.}\ }\textbf {\bibinfo
  {volume} {524}},\ \bibinfo {pages} {012007} (\bibinfo {year}
  {2014})}\BibitemShut {NoStop}%
\bibitem [{dfw()}]{dfwind}%
  \BibitemOpen
  \href {www.dfwind.de} {\bibinfo {title} {www.dfwind.de}}\BibitemShut
  {NoStop}%
\bibitem [{\citenamefont {Neuhaus}\ \emph {et~al.}(2020)\citenamefont
  {Neuhaus}, \citenamefont {H{\"o}lling}, \citenamefont {Bos},\ and\
  \citenamefont {Peinke}}]{neuhaus2020generation}%
  \BibitemOpen
  \bibfield  {author} {\bibinfo {author} {\bibfnamefont {L.}~\bibnamefont
  {Neuhaus}}, \bibinfo {author} {\bibfnamefont {M.}~\bibnamefont
  {H{\"o}lling}}, \bibinfo {author} {\bibfnamefont {W.}~\bibnamefont {Bos}},\
  and\ \bibinfo {author} {\bibfnamefont {J.}~\bibnamefont {Peinke}},\
  }\bibfield  {title} {\bibinfo {title} {Generation of atmospheric turbulence
  with unprecedentedly large {R}eynolds number in a wind tunnel},\ }\href@noop
  {} {\bibfield  {journal} {\bibinfo  {journal} {Phys. Rev. Lett.}\ }\textbf
  {\bibinfo {volume} {125}},\ \bibinfo {pages} {154503} (\bibinfo {year}
  {2020})}\BibitemShut {NoStop}%
\bibitem [{\citenamefont {Cardy}\ \emph {et~al.}(2008)\citenamefont {Cardy},
  \citenamefont {Falkovich},\ and\ \citenamefont {Gawedzki}}]{Cardy_2008}%
  \BibitemOpen
  \bibfield  {author} {\bibinfo {author} {\bibfnamefont {J.}~\bibnamefont
  {Cardy}}, \bibinfo {author} {\bibfnamefont {G.}~\bibnamefont {Falkovich}},\
  and\ \bibinfo {author} {\bibfnamefont {K.}~\bibnamefont {Gawedzki}},\ }\href
  {https://doi.org/10.1017/cbo9780511812149} {\emph {\bibinfo {title}
  {Non-equilibrium Statistical Mechanics and Turbulence}}},\ edited by\
  \bibinfo {editor} {\bibfnamefont {S.}~\bibnamefont {Nazarenko}}\ and\
  \bibinfo {editor} {\bibfnamefont {O.~V.}\ \bibnamefont {Zaboronski}}\
  (\bibinfo  {publisher} {Cambridge University Press},\ \bibinfo {year}
  {2008})\BibitemShut {NoStop}%
\bibitem [{\citenamefont {Beck}\ and\ \citenamefont
  {Cohen}(2003)}]{BECK2003267}%
  \BibitemOpen
  \bibfield  {author} {\bibinfo {author} {\bibfnamefont {C.}~\bibnamefont
  {Beck}}\ and\ \bibinfo {author} {\bibfnamefont {E.}~\bibnamefont {Cohen}},\
  }\bibfield  {title} {\bibinfo {title} {Superstatistics},\ }\href
  {https://doi.org/https://doi.org/10.1016/S0378-4371(03)00019-0} {\bibfield
  {journal} {\bibinfo  {journal} {Physica A}\ }\textbf {\bibinfo {volume}
  {322}},\ \bibinfo {pages} {267} (\bibinfo {year} {2003})}\BibitemShut
  {NoStop}%
\bibitem [{\citenamefont {Metzler}(2020)}]{metzler2020superstatistics}%
  \BibitemOpen
  \bibfield  {author} {\bibinfo {author} {\bibfnamefont {R.}~\bibnamefont
  {Metzler}},\ }\bibfield  {title} {\bibinfo {title} {Superstatistics and
  non-gaussian diffusion},\ }\href@noop {} {\bibfield  {journal} {\bibinfo
  {journal} {Eur. Phys. J.: Spec. Top.}\ }\textbf {\bibinfo {volume} {229}},\
  \bibinfo {pages} {711} (\bibinfo {year} {2020})}\BibitemShut {NoStop}%
\bibitem [{\citenamefont {Veers}\ \emph {et~al.}(2019)\citenamefont {Veers}
  \emph {et~al.}}]{Veers:2019aa}%
  \BibitemOpen
  \bibfield  {author} {\bibinfo {author} {\bibfnamefont {P.}~\bibnamefont
  {Veers}} \emph {et~al.},\ }\bibfield  {title} {\bibinfo {title} {Grand
  challenges in the science of wind energy},\ }\href@noop {} {\bibfield
  {journal} {\bibinfo  {journal} {Science}\ }\textbf {\bibinfo {volume} {366}}
  (\bibinfo {year} {2019})}\BibitemShut {NoStop}%
\bibitem [{\citenamefont {Meneveau}(2019)}]{Meneveau_2019}%
  \BibitemOpen
  \bibfield  {author} {\bibinfo {author} {\bibfnamefont {C.}~\bibnamefont
  {Meneveau}},\ }\bibfield  {title} {\bibinfo {title} {Big wind power: seven
  questions for turbulence research},\ }\href
  {https://doi.org/10.1080/14685248.2019.1584664} {\bibfield  {journal}
  {\bibinfo  {journal} {Journal of Turbulence}\ }\textbf {\bibinfo {volume}
  {20}},\ \bibinfo {pages} {2} (\bibinfo {year} {2019})}\BibitemShut {NoStop}%
\bibitem [{\citenamefont {Mann}(1998)}]{MANN1998269}%
  \BibitemOpen
  \bibfield  {author} {\bibinfo {author} {\bibfnamefont {J.}~\bibnamefont
  {Mann}},\ }\bibfield  {title} {\bibinfo {title} {Wind field simulation},\
  }\href@noop {} {\bibfield  {journal} {\bibinfo  {journal} {Probabilistic Eng.
  Mech.}\ }\textbf {\bibinfo {volume} {13}},\ \bibinfo {pages} {269} (\bibinfo
  {year} {1998})}\BibitemShut {NoStop}%
\bibitem [{\citenamefont {Veers}(1984)}]{veers1984modeling}%
  \BibitemOpen
  \bibfield  {author} {\bibinfo {author} {\bibfnamefont {P.}~\bibnamefont
  {Veers}},\ }\bibfield  {title} {\bibinfo {title} {Modeling stochastic wind
  loads on vertical axis wind turbines},\ }in\ \href@noop {} {\emph {\bibinfo
  {booktitle} {25th Structures, Structural Dynamics and Materials
  Conference}}}\ (\bibinfo {year} {1984})\ p.\ \bibinfo {pages}
  {910}\BibitemShut {NoStop}%
\bibitem [{int(2005)}]{international2005wind}%
  \BibitemOpen
  \href@noop {} {\emph {\bibinfo {title} {Wind Turbines---Part 1: Design
  Requirements: International Standard IEC 61400-1}}} (\bibinfo {year}
  {2005})\BibitemShut {NoStop}%
\bibitem [{\citenamefont {M{\"u}cke}\ \emph {et~al.}(2011)\citenamefont
  {M{\"u}cke}, \citenamefont {Kleinhans},\ and\ \citenamefont
  {Peinke}}]{muecke}%
  \BibitemOpen
  \bibfield  {author} {\bibinfo {author} {\bibfnamefont {T.}~\bibnamefont
  {M{\"u}cke}}, \bibinfo {author} {\bibfnamefont {D.}~\bibnamefont
  {Kleinhans}},\ and\ \bibinfo {author} {\bibfnamefont {J.}~\bibnamefont
  {Peinke}},\ }\bibfield  {title} {\bibinfo {title} {Atmospheric turbulence and
  its influence on the alternating loads on wind turbines},\ }\href@noop {}
  {\bibfield  {journal} {\bibinfo  {journal} {Wind Energy}\ }\textbf {\bibinfo
  {volume} {14}},\ \bibinfo {pages} {301} (\bibinfo {year} {2011})}\BibitemShut
  {NoStop}%
\bibitem [{\citenamefont {Hannesd{\'o}ttir}\ \emph {et~al.}(2019)\citenamefont
  {Hannesd{\'o}ttir}, \citenamefont {Kelly},\ and\ \citenamefont
  {Dimitrov}}]{hannesdottir2019extreme}%
  \BibitemOpen
  \bibfield  {author} {\bibinfo {author} {\bibfnamefont {{\'A}.}~\bibnamefont
  {Hannesd{\'o}ttir}}, \bibinfo {author} {\bibfnamefont {M.}~\bibnamefont
  {Kelly}},\ and\ \bibinfo {author} {\bibfnamefont {N.}~\bibnamefont
  {Dimitrov}},\ }\bibfield  {title} {\bibinfo {title} {Extreme wind
  fluctuations: joint statistics, extreme turbulence, and impact on wind
  turbine loads},\ }\href@noop {} {\bibfield  {journal} {\bibinfo  {journal}
  {Wind Energy Science}\ }\textbf {\bibinfo {volume} {4}},\ \bibinfo {pages}
  {325} (\bibinfo {year} {2019})}\BibitemShut {NoStop}%
\bibitem [{\citenamefont {Gontier}\ \emph {et~al.}(2007)\citenamefont
  {Gontier}, \citenamefont {Schaffarczyk}, \citenamefont {Kleinhans},\ and\
  \citenamefont {Friedrich}}]{Gontier_2007}%
  \BibitemOpen
  \bibfield  {author} {\bibinfo {author} {\bibfnamefont {H.}~\bibnamefont
  {Gontier}}, \bibinfo {author} {\bibfnamefont {A.~P.}\ \bibnamefont
  {Schaffarczyk}}, \bibinfo {author} {\bibfnamefont {D.}~\bibnamefont
  {Kleinhans}},\ and\ \bibinfo {author} {\bibfnamefont {R.}~\bibnamefont
  {Friedrich}},\ }\bibfield  {title} {\bibinfo {title} {A comparison of fatigue
  loads of wind turbine resulting from a non-gaussian turbulence model vs.
  standard ones},\ }\href {https://doi.org/10.1088/1742-6596/75/1/012070}
  {\bibfield  {journal} {\bibinfo  {journal} {J. Phys. Conf. Ser.}\ }\textbf
  {\bibinfo {volume} {75}},\ \bibinfo {pages} {012070} (\bibinfo {year}
  {2007})}\BibitemShut {NoStop}%
\bibitem [{\citenamefont {Kleinhans}\ \emph {et~al.}(2006)\citenamefont
  {Kleinhans}, \citenamefont {Friedrich}, \citenamefont {Gontier},\ and\
  \citenamefont {Schaffarczyk}}]{kleinhans2006simulation}%
  \BibitemOpen
  \bibfield  {author} {\bibinfo {author} {\bibfnamefont {D.}~\bibnamefont
  {Kleinhans}}, \bibinfo {author} {\bibfnamefont {R.}~\bibnamefont
  {Friedrich}}, \bibinfo {author} {\bibfnamefont {H.}~\bibnamefont {Gontier}},\
  and\ \bibinfo {author} {\bibfnamefont {A.}~\bibnamefont {Schaffarczyk}},\
  }\bibfield  {title} {\bibinfo {title} {Simulation of intermittent wind
  fields: A new approach},\ }in\ \href@noop {} {\emph {\bibinfo {booktitle}
  {Proceedings of DEWEK}}},\ Vol.\ \bibinfo {volume} {2006}\ (\bibinfo {year}
  {2006})\BibitemShut {NoStop}%
\bibitem [{\citenamefont {Yassin}\ \emph {et~al.}(2021)\citenamefont {Yassin},
  \citenamefont {Helms}, \citenamefont {Moreno}, \citenamefont {Kassem},
  \citenamefont {H{\"o}ning},\ and\ \citenamefont {Lukassen}}]{Yassin_2021}%
  \BibitemOpen
  \bibfield  {author} {\bibinfo {author} {\bibfnamefont {K.}~\bibnamefont
  {Yassin}}, \bibinfo {author} {\bibfnamefont {A.}~\bibnamefont {Helms}},
  \bibinfo {author} {\bibfnamefont {D.}~\bibnamefont {Moreno}}, \bibinfo
  {author} {\bibfnamefont {H.}~\bibnamefont {Kassem}}, \bibinfo {author}
  {\bibfnamefont {L.}~\bibnamefont {H{\"o}ning}},\ and\ \bibinfo {author}
  {\bibfnamefont {L.~J.}\ \bibnamefont {Lukassen}},\ }\bibfield  {title}
  {\bibinfo {title} {Applying a random time mapping to mann modelled turbulence
  for the generation of intermittent wind fields}\ }\href
  {https://doi.org/10.5194/wes-2021-139} {10.5194/wes-2021-139} (\bibinfo
  {year} {2021})\BibitemShut {NoStop}%
\bibitem [{\citenamefont {Fogedby}(1994)}]{fogedby1994langevin}%
  \BibitemOpen
  \bibfield  {author} {\bibinfo {author} {\bibfnamefont {H.~C.}\ \bibnamefont
  {Fogedby}},\ }\bibfield  {title} {\bibinfo {title} {Langevin equations for
  continuous time l{\'e}vy flights},\ }\href@noop {} {\bibfield  {journal}
  {\bibinfo  {journal} {Phys. Rev. E}\ }\textbf {\bibinfo {volume} {50}},\
  \bibinfo {pages} {1657} (\bibinfo {year} {1994})}\BibitemShut {NoStop}%
\bibitem [{\citenamefont {Eule}\ and\ \citenamefont
  {Friedrich}(2009)}]{Eule_2009}%
  \BibitemOpen
  \bibfield  {author} {\bibinfo {author} {\bibfnamefont {S.}~\bibnamefont
  {Eule}}\ and\ \bibinfo {author} {\bibfnamefont {R.}~\bibnamefont
  {Friedrich}},\ }\bibfield  {title} {\bibinfo {title} {Subordinated langevin
  equations for anomalous diffusion in external potentials {\textemdash}biasing
  and decoupled external forces},\ }\href
  {https://doi.org/10.1209/0295-5075/86/30008} {\bibfield  {journal} {\bibinfo
  {journal} {{EPL}}\ }\textbf {\bibinfo {volume} {86}},\ \bibinfo {pages}
  {30008} (\bibinfo {year} {2009})}\BibitemShut {NoStop}%
\bibitem [{\citenamefont {Kleinhans}(2008)}]{kleinhans2008stochastic}%
  \BibitemOpen
  \bibfield  {author} {\bibinfo {author} {\bibfnamefont {D.}~\bibnamefont
  {Kleinhans}},\ }\bibfield  {title} {\bibinfo {title} {Stochastic modeling of
  complex systems. from the theoretical foundations to the simulation of
  atmospheric wind fields},\ }\href@noop {} {\bibfield  {journal} {\bibinfo
  {journal} {PhD thesis, University of M\"unster}\ } (\bibinfo {year}
  {2008})}\BibitemShut {NoStop}%
\bibitem [{\citenamefont {Beck}\ and\ \citenamefont
  {K{\"u}hn}(2017)}]{Beck_2017}%
  \BibitemOpen
  \bibfield  {author} {\bibinfo {author} {\bibfnamefont {H.}~\bibnamefont
  {Beck}}\ and\ \bibinfo {author} {\bibfnamefont {M.}~\bibnamefont
  {K{\"u}hn}},\ }\bibfield  {title} {\bibinfo {title} {Dynamic data filtering
  of long-range doppler {LiDAR} wind speed measurements},\ }\href@noop {}
  {\bibfield  {journal} {\bibinfo  {journal} {Remote Sensing}\ }\textbf
  {\bibinfo {volume} {9}},\ \bibinfo {pages} {561} (\bibinfo {year}
  {2017})}\BibitemShut {NoStop}%
\bibitem [{\citenamefont {van Dooren}\ \emph {et~al.}(2022)\citenamefont {van
  Dooren}, \citenamefont {Kidambi~Sekar}, \citenamefont {Neuhaus},
  \citenamefont {Mikkelsen}, \citenamefont {H\"olling},\ and\ \citenamefont
  {K\"uhn}}]{amt-15-1355-2022}%
  \BibitemOpen
  \bibfield  {author} {\bibinfo {author} {\bibfnamefont {M.~F.}\ \bibnamefont
  {van Dooren}}, \bibinfo {author} {\bibfnamefont {A.~P.}\ \bibnamefont
  {Kidambi~Sekar}}, \bibinfo {author} {\bibfnamefont {L.}~\bibnamefont
  {Neuhaus}}, \bibinfo {author} {\bibfnamefont {T.}~\bibnamefont {Mikkelsen}},
  \bibinfo {author} {\bibfnamefont {M.}~\bibnamefont {H\"olling}},\ and\
  \bibinfo {author} {\bibfnamefont {M.}~\bibnamefont {K\"uhn}},\ }\bibfield
  {title} {\bibinfo {title} {Modelling the spectral shape of continuous-wave
  lidar measurements in a turbulent wind tunnel},\ }\href
  {https://doi.org/10.5194/amt-15-1355-2022} {\bibfield  {journal} {\bibinfo
  {journal} {Atmos. Meas. Tech.}\ }\textbf {\bibinfo {volume} {15}},\ \bibinfo
  {pages} {1355} (\bibinfo {year} {2022})}\BibitemShut {NoStop}%
\bibitem [{x_w()}]{x_wakes}%
  \BibitemOpen
  \href@noop {} {\bibinfo {title}
  {https://magazin.tu-braunschweig.de/pi-post/messung-von-windparkeffekten-ueber-der-nordsee-mit-zwei-flugzeugen/}}\BibitemShut
  {NoStop}%
\bibitem [{\citenamefont {Buccolieri}\ \emph {et~al.}(2015)\citenamefont
  {Buccolieri}, \citenamefont {Salizzoni}, \citenamefont {Soulhac},
  \citenamefont {Garbero},\ and\ \citenamefont
  {Di~Sabatino}}]{buccolieri2015breathability}%
  \BibitemOpen
  \bibfield  {author} {\bibinfo {author} {\bibfnamefont {R.}~\bibnamefont
  {Buccolieri}}, \bibinfo {author} {\bibfnamefont {P.}~\bibnamefont
  {Salizzoni}}, \bibinfo {author} {\bibfnamefont {L.}~\bibnamefont {Soulhac}},
  \bibinfo {author} {\bibfnamefont {V.}~\bibnamefont {Garbero}},\ and\ \bibinfo
  {author} {\bibfnamefont {S.}~\bibnamefont {Di~Sabatino}},\ }\bibfield
  {title} {\bibinfo {title} {The breathability of compact cities},\ }\href@noop
  {} {\bibfield  {journal} {\bibinfo  {journal} {Urban Climate}\ }\textbf
  {\bibinfo {volume} {13}},\ \bibinfo {pages} {73} (\bibinfo {year}
  {2015})}\BibitemShut {NoStop}%
\bibitem [{\citenamefont {Salizzoni}\ \emph {et~al.}(2011)\citenamefont
  {Salizzoni}, \citenamefont {Marro}, \citenamefont {Soulhac}, \citenamefont
  {Grosjean},\ and\ \citenamefont {Perkins}}]{salizzoni2011turbulent}%
  \BibitemOpen
  \bibfield  {author} {\bibinfo {author} {\bibfnamefont {P.}~\bibnamefont
  {Salizzoni}}, \bibinfo {author} {\bibfnamefont {M.}~\bibnamefont {Marro}},
  \bibinfo {author} {\bibfnamefont {L.}~\bibnamefont {Soulhac}}, \bibinfo
  {author} {\bibfnamefont {N.}~\bibnamefont {Grosjean}},\ and\ \bibinfo
  {author} {\bibfnamefont {R.~J.}\ \bibnamefont {Perkins}},\ }\bibfield
  {title} {\bibinfo {title} {Turbulent transfer between street canyons and the
  overlying atmospheric boundary layer},\ }\href@noop {} {\bibfield  {journal}
  {\bibinfo  {journal} {Bound.-Layer Meteorol.}\ }\textbf {\bibinfo {volume}
  {141}},\ \bibinfo {pages} {393} (\bibinfo {year} {2011})}\BibitemShut
  {NoStop}%
\bibitem [{\citenamefont {Voelkel}\ and\ \citenamefont
  {Shandas}(2017)}]{voelkel2017towards}%
  \BibitemOpen
  \bibfield  {author} {\bibinfo {author} {\bibfnamefont {J.}~\bibnamefont
  {Voelkel}}\ and\ \bibinfo {author} {\bibfnamefont {V.}~\bibnamefont
  {Shandas}},\ }\bibfield  {title} {\bibinfo {title} {Towards systematic
  prediction of urban heat islands: Grounding measurements, assessing modeling
  techniques},\ }\href@noop {} {\bibfield  {journal} {\bibinfo  {journal}
  {Climate}\ }\textbf {\bibinfo {volume} {5}},\ \bibinfo {pages} {41} (\bibinfo
  {year} {2017})}\BibitemShut {NoStop}%
\bibitem [{\citenamefont {Cassiani}\ \emph {et~al.}(2010)\citenamefont
  {Cassiani}, \citenamefont {Vinuesa}, \citenamefont {Galmarini},\ and\
  \citenamefont {Denby}}]{cassiani2010stochastic}%
  \BibitemOpen
  \bibfield  {author} {\bibinfo {author} {\bibfnamefont {M.}~\bibnamefont
  {Cassiani}}, \bibinfo {author} {\bibfnamefont {J.-F.}\ \bibnamefont
  {Vinuesa}}, \bibinfo {author} {\bibfnamefont {S.}~\bibnamefont {Galmarini}},\
  and\ \bibinfo {author} {\bibfnamefont {B.}~\bibnamefont {Denby}},\ }\bibfield
   {title} {\bibinfo {title} {Stochastic fields method for sub-grid scale
  emission heterogeneity in mesoscale atmospheric dispersion models},\
  }\href@noop {} {\bibfield  {journal} {\bibinfo  {journal} {Atmos. Chem.
  Phys.}\ }\textbf {\bibinfo {volume} {10}},\ \bibinfo {pages} {267} (\bibinfo
  {year} {2010})}\BibitemShut {NoStop}%
\bibitem [{\citenamefont {Lengyel}\ \emph {et~al.}(2021)\citenamefont
  {Lengyel}, \citenamefont {Alvenides},\ and\ \citenamefont {Friedrich}}]{epb}%
  \BibitemOpen
  \bibfield  {author} {\bibinfo {author} {\bibfnamefont {J.}~\bibnamefont
  {Lengyel}}, \bibinfo {author} {\bibfnamefont {S.}~\bibnamefont {Alvenides}},\
  and\ \bibinfo {author} {\bibfnamefont {J.}~\bibnamefont {Friedrich}},\
  }\bibfield  {title} {\bibinfo {title} {Modelling the interdependence of
  spatial scales in urban systems},\ }\href@noop {} {\bibfield  {journal}
  {\bibinfo  {journal} {to appear in Environ. Plann. B}\ } (\bibinfo {year}
  {2021})}\BibitemShut {NoStop}%
\bibitem [{\citenamefont {Koerber}\ \emph {et~al.}(1988)\citenamefont
  {Koerber}, \citenamefont {Besel},\ and\ \citenamefont
  {Reinhold}}]{koerber19883}%
  \BibitemOpen
  \bibfield  {author} {\bibinfo {author} {\bibfnamefont {F.}~\bibnamefont
  {Koerber}}, \bibinfo {author} {\bibfnamefont {G.}~\bibnamefont {Besel}},\
  and\ \bibinfo {author} {\bibfnamefont {H.}~\bibnamefont {Reinhold}},\
  }\bibfield  {title} {\bibinfo {title} {3 {MW} {GROWIAN} wind turbine test
  program. {F}inal report. {M}essprogramm an der 3 {MW}-{W}indkraftanlage
  {GROWIAN}. {S}chlussbericht},\ }\href@noop {} {\  (\bibinfo {year}
  {1988})}\BibitemShut {NoStop}%
\bibitem [{\citenamefont {G\"unther}\ and\ \citenamefont
  {Hennemuth}(1998)}]{dwd}%
  \BibitemOpen
  \bibfield  {author} {\bibinfo {author} {\bibfnamefont {H.}~\bibnamefont
  {G\"unther}}\ and\ \bibinfo {author} {\bibfnamefont {B.}~\bibnamefont
  {Hennemuth}},\ }\bibfield  {title} {\bibinfo {title} {{E}rste {A}ufbereitung
  von fl{\"a}chenhaften {W}indmessdaten in {H}{\"o}hen bis 150m},\ }\href@noop
  {} {\bibfield  {journal} {\bibinfo  {journal} {Deutscher Wetter Dienst}\
  }\textbf {\bibinfo {volume} {BMBF-Projekt}},\ \bibinfo {pages} {0329372A}
  (\bibinfo {year} {1998})}\BibitemShut {NoStop}%
\bibitem [{\citenamefont {Winter}(2016)}]{winter2016modellierung}%
  \BibitemOpen
  \bibfield  {author} {\bibinfo {author} {\bibfnamefont {T.~A.}\ \bibnamefont
  {Winter}},\ }\emph {\bibinfo {title} {Modellierung dynamischer Lasten auf
  Windkraftanlagen unter dem Einfluss turbulenter Anstr{\"o}mung}},\ \href@noop
  {} {Ph.D. thesis},\ \bibinfo  {school} {Carl von Ossietzky Universit{\"a}t
  Oldenburg} (\bibinfo {year} {2016})\BibitemShut {NoStop}%
\bibitem [{\citenamefont {Horst}(1973)}]{horst1973corrections}%
  \BibitemOpen
  \bibfield  {author} {\bibinfo {author} {\bibfnamefont {T.~W.}\ \bibnamefont
  {Horst}},\ }\bibfield  {title} {\bibinfo {title} {Corrections for response
  errors in a three-component propeller anemometer},\ }\href@noop {} {\bibfield
   {journal} {\bibinfo  {journal} {Journal of Applied Meteorology and
  Climatology}\ }\textbf {\bibinfo {volume} {12}},\ \bibinfo {pages} {716}
  (\bibinfo {year} {1973})}\BibitemShut {NoStop}%
\bibitem [{\citenamefont {Veers}(1988)}]{veers1988three}%
  \BibitemOpen
  \bibfield  {author} {\bibinfo {author} {\bibfnamefont {P.~S.}\ \bibnamefont
  {Veers}},\ }\href@noop {} {\emph {\bibinfo {title} {Three-dimensional wind
  simulation}}},\ \bibinfo {type} {Tech. Rep.}\ (\bibinfo  {institution}
  {Sandia National Labs., Albuquerque, NM (USA)},\ \bibinfo {year}
  {1988})\BibitemShut {NoStop}%
\bibitem [{\citenamefont {Frisch}(1995)}]{frisch:1995}%
  \BibitemOpen
  \bibfield  {author} {\bibinfo {author} {\bibfnamefont {U.}~\bibnamefont
  {Frisch}},\ }\href@noop {} {\emph {\bibinfo {title} {{Turbulence}}}}\
  (\bibinfo  {publisher} {Cambridge University Press},\ \bibinfo {year}
  {1995})\BibitemShut {NoStop}%
\bibitem [{\citenamefont {Taylor}(1935)}]{Taylor1935}%
  \BibitemOpen
  \bibfield  {author} {\bibinfo {author} {\bibfnamefont {G.~I.}\ \bibnamefont
  {Taylor}},\ }\bibfield  {title} {\bibinfo {title} {{Statistical Theory of
  Turbulence}},\ }\href {https://doi.org/10.1098/rspa.1935.0158} {\bibfield
  {journal} {\bibinfo  {journal} {Proc. R. Soc. London A Math. Phys. Eng.
  Sci.}\ }\textbf {\bibinfo {volume} {151}},\ \bibinfo {pages} {421} (\bibinfo
  {year} {1935})}\BibitemShut {NoStop}%
\bibitem [{\citenamefont {Friedrich}\ and\ \citenamefont
  {Peinke}(1997)}]{Friedrich1997}%
  \BibitemOpen
  \bibfield  {author} {\bibinfo {author} {\bibfnamefont {R.}~\bibnamefont
  {Friedrich}}\ and\ \bibinfo {author} {\bibfnamefont {J.}~\bibnamefont
  {Peinke}},\ }\bibfield  {title} {\bibinfo {title} {{Description of a
  Turbulent Cascade by a Fokker-Planck Equation}},\ }\href@noop {} {\bibfield
  {journal} {\bibinfo  {journal} {Phys. Rev. Lett.}\ }\textbf {\bibinfo
  {volume} {78}},\ \bibinfo {pages} {863} (\bibinfo {year} {1997})}\BibitemShut
  {NoStop}%
\bibitem [{\citenamefont {Boettcher}\ \emph {et~al.}(2003)\citenamefont
  {Boettcher}, \citenamefont {Renner}, \citenamefont {Waldl},\ and\
  \citenamefont {Peinke}}]{boettcher2003statistics}%
  \BibitemOpen
  \bibfield  {author} {\bibinfo {author} {\bibfnamefont {F.}~\bibnamefont
  {Boettcher}}, \bibinfo {author} {\bibfnamefont {C.}~\bibnamefont {Renner}},
  \bibinfo {author} {\bibfnamefont {H.-P.}\ \bibnamefont {Waldl}},\ and\
  \bibinfo {author} {\bibfnamefont {J.}~\bibnamefont {Peinke}},\ }\bibfield
  {title} {\bibinfo {title} {On the statistics of wind gusts},\ }\href@noop {}
  {\bibfield  {journal} {\bibinfo  {journal} {Bound.-Layer Meteorol.}\ }\textbf
  {\bibinfo {volume} {108}},\ \bibinfo {pages} {163} (\bibinfo {year}
  {2003})}\BibitemShut {NoStop}%
\bibitem [{\citenamefont {Juneja}\ \emph {et~al.}(1994)\citenamefont {Juneja},
  \citenamefont {Lathrop}, \citenamefont {Sreenivasan},\ and\ \citenamefont
  {Stolovitzky}}]{juneja1994synthetic}%
  \BibitemOpen
  \bibfield  {author} {\bibinfo {author} {\bibfnamefont {A.}~\bibnamefont
  {Juneja}}, \bibinfo {author} {\bibfnamefont {D.~P.}\ \bibnamefont {Lathrop}},
  \bibinfo {author} {\bibfnamefont {K.~R.}\ \bibnamefont {Sreenivasan}},\ and\
  \bibinfo {author} {\bibfnamefont {G.}~\bibnamefont {Stolovitzky}},\
  }\bibfield  {title} {\bibinfo {title} {Synthetic turbulence},\ }\href@noop {}
  {\bibfield  {journal} {\bibinfo  {journal} {Phys. Rev. E}\ }\textbf {\bibinfo
  {volume} {49}},\ \bibinfo {pages} {5179} (\bibinfo {year}
  {1994})}\BibitemShut {NoStop}%
\bibitem [{\citenamefont {Malara}\ \emph {et~al.}(2016)\citenamefont {Malara},
  \citenamefont {Di~Mare}, \citenamefont {Nigro},\ and\ \citenamefont
  {Sorriso-Valvo}}]{malara2016fast}%
  \BibitemOpen
  \bibfield  {author} {\bibinfo {author} {\bibfnamefont {F.}~\bibnamefont
  {Malara}}, \bibinfo {author} {\bibfnamefont {F.}~\bibnamefont {Di~Mare}},
  \bibinfo {author} {\bibfnamefont {G.}~\bibnamefont {Nigro}},\ and\ \bibinfo
  {author} {\bibfnamefont {L.}~\bibnamefont {Sorriso-Valvo}},\ }\bibfield
  {title} {\bibinfo {title} {Fast algorithm for a three-dimensional synthetic
  model of intermittent turbulence},\ }\href@noop {} {\bibfield  {journal}
  {\bibinfo  {journal} {Phys. Rev. E}\ }\textbf {\bibinfo {volume} {94}},\
  \bibinfo {pages} {053109} (\bibinfo {year} {2016})}\BibitemShut {NoStop}%
\bibitem [{\citenamefont {Rosales}\ and\ \citenamefont
  {Meneveau}(2008)}]{rosales2008anomalous}%
  \BibitemOpen
  \bibfield  {author} {\bibinfo {author} {\bibfnamefont {C.}~\bibnamefont
  {Rosales}}\ and\ \bibinfo {author} {\bibfnamefont {C.}~\bibnamefont
  {Meneveau}},\ }\bibfield  {title} {\bibinfo {title} {Anomalous scaling and
  intermittency in three-dimensional synthetic turbulence},\ }\href@noop {}
  {\bibfield  {journal} {\bibinfo  {journal} {Phys. Rev. E}\ }\textbf {\bibinfo
  {volume} {78}},\ \bibinfo {pages} {016313} (\bibinfo {year}
  {2008})}\BibitemShut {NoStop}%
\bibitem [{\citenamefont {Lovejoy}\ and\ \citenamefont
  {Schertzer}(1986)}]{lovejoy1986scale}%
  \BibitemOpen
  \bibfield  {author} {\bibinfo {author} {\bibfnamefont {S.}~\bibnamefont
  {Lovejoy}}\ and\ \bibinfo {author} {\bibfnamefont {D.}~\bibnamefont
  {Schertzer}},\ }\bibfield  {title} {\bibinfo {title} {Scale invariance,
  symmetries, fractals, and stochastic simulations of atmospheric phenomena},\
  }\href@noop {} {\bibfield  {journal} {\bibinfo  {journal} {Bull. Am. Met.
  Soc.}\ }\textbf {\bibinfo {volume} {67}},\ \bibinfo {pages} {21} (\bibinfo
  {year} {1986})}\BibitemShut {NoStop}%
\bibitem [{\citenamefont {Bacry}\ \emph {et~al.}(2001)\citenamefont {Bacry},
  \citenamefont {Delour},\ and\ \citenamefont {Muzy}}]{bacry2001multifractal}%
  \BibitemOpen
  \bibfield  {author} {\bibinfo {author} {\bibfnamefont {E.}~\bibnamefont
  {Bacry}}, \bibinfo {author} {\bibfnamefont {J.}~\bibnamefont {Delour}},\ and\
  \bibinfo {author} {\bibfnamefont {J.-F.}\ \bibnamefont {Muzy}},\ }\bibfield
  {title} {\bibinfo {title} {Multifractal random walk},\ }\href@noop {}
  {\bibfield  {journal} {\bibinfo  {journal} {Phys. Rev. E}\ }\textbf {\bibinfo
  {volume} {64}},\ \bibinfo {pages} {026103} (\bibinfo {year}
  {2001})}\BibitemShut {NoStop}%
\bibitem [{\citenamefont {Chevillard}\ \emph {et~al.}(2010)\citenamefont
  {Chevillard}, \citenamefont {Robert},\ and\ \citenamefont
  {Vargas}}]{chevillard2010stochastic}%
  \BibitemOpen
  \bibfield  {author} {\bibinfo {author} {\bibfnamefont {L.}~\bibnamefont
  {Chevillard}}, \bibinfo {author} {\bibfnamefont {R.}~\bibnamefont {Robert}},\
  and\ \bibinfo {author} {\bibfnamefont {V.}~\bibnamefont {Vargas}},\
  }\bibfield  {title} {\bibinfo {title} {A stochastic representation of the
  local structure of turbulence},\ }\href@noop {} {\bibfield  {journal}
  {\bibinfo  {journal} {EPL}\ }\textbf {\bibinfo {volume} {89}},\ \bibinfo
  {pages} {54002} (\bibinfo {year} {2010})}\BibitemShut {NoStop}%
\bibitem [{\citenamefont {Friedrich}\ \emph {et~al.}(2021)\citenamefont
  {Friedrich}, \citenamefont {Peinke}, \citenamefont {Pumir},\ and\
  \citenamefont {Grauer}}]{Friedrich_2021}%
  \BibitemOpen
  \bibfield  {author} {\bibinfo {author} {\bibfnamefont {J.}~\bibnamefont
  {Friedrich}}, \bibinfo {author} {\bibfnamefont {J.}~\bibnamefont {Peinke}},
  \bibinfo {author} {\bibfnamefont {A.}~\bibnamefont {Pumir}},\ and\ \bibinfo
  {author} {\bibfnamefont {R.}~\bibnamefont {Grauer}},\ }\bibfield  {title}
  {\bibinfo {title} {Explicit construction of joint~multipoint~statistics in
  complex systems},\ }\href@noop {} {\bibfield  {journal} {\bibinfo  {journal}
  {J. Phys. Complexity}\ }\textbf {\bibinfo {volume} {2}},\ \bibinfo {pages}
  {045006} (\bibinfo {year} {2021})}\BibitemShut {NoStop}%
\bibitem [{\citenamefont {Kolmogorov}(1962)}]{Kolmogorov1962}%
  \BibitemOpen
  \bibfield  {author} {\bibinfo {author} {\bibfnamefont {A.~N.}\ \bibnamefont
  {Kolmogorov}},\ }\bibfield  {title} {\bibinfo {title} {{A refinement of
  previous hypotheses concerning the local structure of turbulence in a viscous
  incompressible fluid at high Reynolds number}},\ }\href@noop {} {\bibfield
  {journal} {\bibinfo  {journal} {J. Fluid Mech.}\ }\textbf {\bibinfo {volume}
  {13}},\ \bibinfo {pages} {82} (\bibinfo {year} {1962})}\BibitemShut {NoStop}%
\bibitem [{\citenamefont {Kolmogorov}(1941)}]{Kolmogorov1941}%
  \BibitemOpen
  \bibfield  {author} {\bibinfo {author} {\bibfnamefont {A.~N.}\ \bibnamefont
  {Kolmogorov}},\ }\bibfield  {title} {\bibinfo {title} {{The local structure
  of turbulence in incompressible viscous fluid for very large Reynolds
  numbers}},\ }\href {https://doi.org/10.1098/rspa.1991.0075} {\bibfield
  {journal} {\bibinfo  {journal} {Dokl. Akad. Nauk Sssr}\ }\textbf {\bibinfo
  {volume} {30}},\ \bibinfo {pages} {301} (\bibinfo {year} {1941})}\BibitemShut
  {NoStop}%
\bibitem [{\citenamefont {Beck}(2007)}]{beck2007statistics}%
  \BibitemOpen
  \bibfield  {author} {\bibinfo {author} {\bibfnamefont {C.}~\bibnamefont
  {Beck}},\ }\bibfield  {title} {\bibinfo {title} {Statistics of
  three-dimensional {L}agrangian turbulence},\ }\href@noop {} {\bibfield
  {journal} {\bibinfo  {journal} {Phys. Rev. Lett.}\ }\textbf {\bibinfo
  {volume} {98}},\ \bibinfo {pages} {064502} (\bibinfo {year}
  {2007})}\BibitemShut {NoStop}%
\bibitem [{\citenamefont {Castaing}\ \emph {et~al.}(1990)\citenamefont
  {Castaing}, \citenamefont {Gagne},\ and\ \citenamefont
  {Hopfinger}}]{CASTAING1990177}%
  \BibitemOpen
  \bibfield  {author} {\bibinfo {author} {\bibfnamefont {B.}~\bibnamefont
  {Castaing}}, \bibinfo {author} {\bibfnamefont {Y.}~\bibnamefont {Gagne}},\
  and\ \bibinfo {author} {\bibfnamefont {E.}~\bibnamefont {Hopfinger}},\
  }\bibfield  {title} {\bibinfo {title} {Velocity probability density functions
  of high reynolds number turbulence},\ }\href
  {https://doi.org/https://doi.org/10.1016/0167-2789(90)90035-N} {\bibfield
  {journal} {\bibinfo  {journal} {Physica D}\ }\textbf {\bibinfo {volume}
  {46}},\ \bibinfo {pages} {177} (\bibinfo {year} {1990})}\BibitemShut
  {NoStop}%
\bibitem [{\citenamefont {Wilczek}(2016)}]{Wilczek_2016}%
  \BibitemOpen
  \bibfield  {author} {\bibinfo {author} {\bibfnamefont {M.}~\bibnamefont
  {Wilczek}},\ }\bibfield  {title} {\bibinfo {title} {Non-{G}aussianity and
  intermittency in an ensemble of {G}aussian fields},\ }\href@noop {}
  {\bibfield  {journal} {\bibinfo  {journal} {New J. Phys.}\ }\textbf {\bibinfo
  {volume} {18}},\ \bibinfo {pages} {125009} (\bibinfo {year}
  {2016})}\BibitemShut {NoStop}%
\bibitem [{\citenamefont {Yakhot}(2006)}]{YAKHOT2006166}%
  \BibitemOpen
  \bibfield  {author} {\bibinfo {author} {\bibfnamefont {V.}~\bibnamefont
  {Yakhot}},\ }\bibfield  {title} {\bibinfo {title} {Probability densities in
  strong turbulence},\ }\href@noop {} {\bibfield  {journal} {\bibinfo
  {journal} {Physica D}\ }\textbf {\bibinfo {volume} {215}},\ \bibinfo {pages}
  {166} (\bibinfo {year} {2006})}\BibitemShut {NoStop}%
\bibitem [{\citenamefont {Monin}\ and\ \citenamefont {Yaglom}(2007)}]{monin}%
  \BibitemOpen
  \bibfield  {author} {\bibinfo {author} {\bibfnamefont {A.~S.}\ \bibnamefont
  {Monin}}\ and\ \bibinfo {author} {\bibfnamefont {A.~M.}\ \bibnamefont
  {Yaglom}},\ }\href@noop {} {\emph {\bibinfo {title} {{Statistical Fluid
  Mechanics: Mechanics of Turbulence}}}}\ (\bibinfo  {publisher} {Courier Dover
  Publications},\ \bibinfo {year} {2007})\BibitemShut {NoStop}%
\bibitem [{\citenamefont {Beck}\ \emph {et~al.}(2005)\citenamefont {Beck},
  \citenamefont {Cohen},\ and\ \citenamefont {Swinney}}]{Beck:2005vx}%
  \BibitemOpen
  \bibfield  {author} {\bibinfo {author} {\bibfnamefont {C.}~\bibnamefont
  {Beck}}, \bibinfo {author} {\bibfnamefont {E.~G.}\ \bibnamefont {Cohen}},\
  and\ \bibinfo {author} {\bibfnamefont {H.~L.}\ \bibnamefont {Swinney}},\
  }\bibfield  {title} {\bibinfo {title} {From time series to superstatistics},\
  }\href@noop {} {\bibfield  {journal} {\bibinfo  {journal} {Phys. Rev. E}\
  }\textbf {\bibinfo {volume} {72}} (\bibinfo {year} {2005})}\BibitemShut
  {NoStop}%
\bibitem [{\citenamefont {Friedrich}(2020)}]{friedrich2020non}%
  \BibitemOpen
  \bibfield  {author} {\bibinfo {author} {\bibfnamefont {J.}~\bibnamefont
  {Friedrich}},\ }\href@noop {} {\emph {\bibinfo {title} {Non-perturbative
  Methods in Statistical Descriptions of Turbulence}}}\ (\bibinfo  {publisher}
  {Springer},\ \bibinfo {year} {2020})\BibitemShut {NoStop}%
\bibitem [{\citenamefont {Mardoukhi}\ \emph {et~al.}(2020)\citenamefont
  {Mardoukhi}, \citenamefont {Chechkin},\ and\ \citenamefont
  {Metzler}}]{Mardoukhi_2020}%
  \BibitemOpen
  \bibfield  {author} {\bibinfo {author} {\bibfnamefont {Y.}~\bibnamefont
  {Mardoukhi}}, \bibinfo {author} {\bibfnamefont {A.}~\bibnamefont
  {Chechkin}},\ and\ \bibinfo {author} {\bibfnamefont {R.}~\bibnamefont
  {Metzler}},\ }\bibfield  {title} {\bibinfo {title} {Spurious ergodicity
  breaking in normal and fractional {O}rnstein{\textendash}{U}hlenbeck
  process},\ }\href@noop {} {\bibfield  {journal} {\bibinfo  {journal} {New J.
  Phys.}\ }\textbf {\bibinfo {volume} {22}},\ \bibinfo {pages} {073012}
  (\bibinfo {year} {2020})}\BibitemShut {NoStop}%
\bibitem [{\citenamefont {Shen}\ and\ \citenamefont
  {Warhaft}(2002)}]{Shen_2002}%
  \BibitemOpen
  \bibfield  {author} {\bibinfo {author} {\bibfnamefont {X.}~\bibnamefont
  {Shen}}\ and\ \bibinfo {author} {\bibfnamefont {Z.}~\bibnamefont {Warhaft}},\
  }\bibfield  {title} {\bibinfo {title} {Longitudinal and transverse structure
  functions in sheared and unsheared wind-tunnel turbulence},\ }\href
  {https://doi.org/10.1063/1.1421059} {\bibfield  {journal} {\bibinfo
  {journal} {Phys. Fluids}\ }\textbf {\bibinfo {volume} {14}},\ \bibinfo
  {pages} {370} (\bibinfo {year} {2002})}\BibitemShut {NoStop}%
\bibitem [{\citenamefont {Friedrich}\ \emph {et~al.}(2020)\citenamefont
  {Friedrich}, \citenamefont {Gallon}, \citenamefont {Pumir},\ and\
  \citenamefont {Grauer}}]{friedrich2020stochastic}%
  \BibitemOpen
  \bibfield  {author} {\bibinfo {author} {\bibfnamefont {J.}~\bibnamefont
  {Friedrich}}, \bibinfo {author} {\bibfnamefont {S.}~\bibnamefont {Gallon}},
  \bibinfo {author} {\bibfnamefont {A.}~\bibnamefont {Pumir}},\ and\ \bibinfo
  {author} {\bibfnamefont {R.}~\bibnamefont {Grauer}},\ }\bibfield  {title}
  {\bibinfo {title} {Stochastic interpolation of sparsely sampled time series
  via multipoint fractional {B}rownian bridges},\ }\href@noop {} {\bibfield
  {journal} {\bibinfo  {journal} {Phys. Rev. Lett.}\ }\textbf {\bibinfo
  {volume} {125}},\ \bibinfo {pages} {170602} (\bibinfo {year}
  {2020})}\BibitemShut {NoStop}%
\bibitem [{\citenamefont {Bierbooms}(2009)}]{bierbooms2009constrained}%
  \BibitemOpen
  \bibfield  {author} {\bibinfo {author} {\bibfnamefont {W.~A.}\ \bibnamefont
  {Bierbooms}},\ }\emph {\bibinfo {title} {Constrained stochastic simulation of
  wind gusts for wind turbine design}},\ \href@noop {} {Ph.D. thesis},\
  \bibinfo  {school} {TU Delft} (\bibinfo {year} {2009})\BibitemShut {NoStop}%
\bibitem [{\citenamefont {Nazarenko}\ and\ \citenamefont
  {Schekochihin}(2011)}]{nazarenko_schekochihin_2011}%
  \BibitemOpen
  \bibfield  {author} {\bibinfo {author} {\bibfnamefont {S.~V.}\ \bibnamefont
  {Nazarenko}}\ and\ \bibinfo {author} {\bibfnamefont {A.~A.}\ \bibnamefont
  {Schekochihin}},\ }\bibfield  {title} {\bibinfo {title} {Critical balance in
  magnetohydrodynamic, rotating and stratified turbulence: towards a universal
  scaling conjecture},\ }\href@noop {} {\bibfield  {journal} {\bibinfo
  {journal} {J. Fluid Mech.}\ }\textbf {\bibinfo {volume} {677}},\ \bibinfo
  {pages} {134} (\bibinfo {year} {2011})}\BibitemShut {NoStop}%
\bibitem [{\citenamefont {Chougule}\ \emph {et~al.}(2017)\citenamefont
  {Chougule}, \citenamefont {Mann}, \citenamefont {Kelly},\ and\ \citenamefont
  {Larsen}}]{chougule2017modeling}%
  \BibitemOpen
  \bibfield  {author} {\bibinfo {author} {\bibfnamefont {A.}~\bibnamefont
  {Chougule}}, \bibinfo {author} {\bibfnamefont {J.}~\bibnamefont {Mann}},
  \bibinfo {author} {\bibfnamefont {M.}~\bibnamefont {Kelly}},\ and\ \bibinfo
  {author} {\bibfnamefont {G.~C.}\ \bibnamefont {Larsen}},\ }\bibfield  {title}
  {\bibinfo {title} {Modeling atmospheric turbulence via rapid distortion
  theory: spectral tensor of velocity and buoyancy},\ }\href@noop {} {\bibfield
   {journal} {\bibinfo  {journal} {J. Atmos. Sci.}\ }\textbf {\bibinfo {volume}
  {74}},\ \bibinfo {pages} {949} (\bibinfo {year} {2017})}\BibitemShut
  {NoStop}%
\bibitem [{\citenamefont {Chougule}\ \emph {et~al.}(2018)\citenamefont
  {Chougule}, \citenamefont {Mann}, \citenamefont {Kelly},\ and\ \citenamefont
  {Larsen}}]{chougule2018simplification}%
  \BibitemOpen
  \bibfield  {author} {\bibinfo {author} {\bibfnamefont {A.}~\bibnamefont
  {Chougule}}, \bibinfo {author} {\bibfnamefont {J.}~\bibnamefont {Mann}},
  \bibinfo {author} {\bibfnamefont {M.}~\bibnamefont {Kelly}},\ and\ \bibinfo
  {author} {\bibfnamefont {G.~C.}\ \bibnamefont {Larsen}},\ }\bibfield  {title}
  {\bibinfo {title} {Simplification and validation of a spectral-tensor model
  for turbulence including atmospheric stability},\ }\href@noop {} {\bibfield
  {journal} {\bibinfo  {journal} {Bound.-Layer Meteorol.}\ }\textbf {\bibinfo
  {volume} {167}},\ \bibinfo {pages} {371} (\bibinfo {year}
  {2018})}\BibitemShut {NoStop}%
\bibitem [{\citenamefont {Chandrasekhar}(1950)}]{1950}%
  \BibitemOpen
  \bibfield  {author} {\bibinfo {author} {\bibfnamefont {S.}~\bibnamefont
  {Chandrasekhar}},\ }\bibfield  {title} {\bibinfo {title} {The theory of
  axisymmetric turbulence},\ }\href@noop {} {\bibfield  {journal} {\bibinfo
  {journal} {Phil. Trans. R. Soc. A}\ }\textbf {\bibinfo {volume} {242}},\
  \bibinfo {pages} {557} (\bibinfo {year} {1950})}\BibitemShut {NoStop}%
\bibitem [{\citenamefont {Robertson}(1940)}]{Robertson1940}%
  \BibitemOpen
  \bibfield  {author} {\bibinfo {author} {\bibfnamefont {H.~P.}\ \bibnamefont
  {Robertson}},\ }\bibfield  {title} {\bibinfo {title} {{The invariant theory
  of isotropic turbulence}},\ }in\ \href@noop {} {\emph {\bibinfo {booktitle}
  {Math. Proc. Cambridge Philos. Soc.}}},\ Vol.~\bibinfo {volume} {36}\
  (\bibinfo  {publisher} {Cambridge Univ Press},\ \bibinfo {year} {1940})\ pp.\
  \bibinfo {pages} {209--223}\BibitemShut {NoStop}%
\bibitem [{\citenamefont {Cheng}\ \emph {et~al.}(2017)\citenamefont {Cheng},
  \citenamefont {Sayde}, \citenamefont {Li}, \citenamefont {Basara},
  \citenamefont {Selker}, \citenamefont {Tanner},\ and\ \citenamefont
  {Gentine}}]{cheng2017failure}%
  \BibitemOpen
  \bibfield  {author} {\bibinfo {author} {\bibfnamefont {Y.}~\bibnamefont
  {Cheng}}, \bibinfo {author} {\bibfnamefont {C.}~\bibnamefont {Sayde}},
  \bibinfo {author} {\bibfnamefont {Q.}~\bibnamefont {Li}}, \bibinfo {author}
  {\bibfnamefont {J.}~\bibnamefont {Basara}}, \bibinfo {author} {\bibfnamefont
  {J.}~\bibnamefont {Selker}}, \bibinfo {author} {\bibfnamefont
  {E.}~\bibnamefont {Tanner}},\ and\ \bibinfo {author} {\bibfnamefont
  {P.}~\bibnamefont {Gentine}},\ }\bibfield  {title} {\bibinfo {title} {Failure
  of taylor's hypothesis in the atmospheric surface layer and its correction
  for eddy-covariance measurements},\ }\href@noop {} {\bibfield  {journal}
  {\bibinfo  {journal} {Geophysical Research Letters}\ }\textbf {\bibinfo
  {volume} {44}},\ \bibinfo {pages} {4287} (\bibinfo {year}
  {2017})}\BibitemShut {NoStop}%
\bibitem [{\citenamefont {Del~Alamo}\ and\ \citenamefont
  {Jim{\'e}nez}(2009)}]{del2009estimation}%
  \BibitemOpen
  \bibfield  {author} {\bibinfo {author} {\bibfnamefont {J.~C.}\ \bibnamefont
  {Del~Alamo}}\ and\ \bibinfo {author} {\bibfnamefont {J.}~\bibnamefont
  {Jim{\'e}nez}},\ }\bibfield  {title} {\bibinfo {title} {Estimation of
  turbulent convection velocities and corrections to taylor's approximation},\
  }\href@noop {} {\bibfield  {journal} {\bibinfo  {journal} {J. Fluid Mech.}\
  }\textbf {\bibinfo {volume} {640}},\ \bibinfo {pages} {5} (\bibinfo {year}
  {2009})}\BibitemShut {NoStop}%
\bibitem [{\citenamefont {Yang}\ and\ \citenamefont
  {Howland}(2018)}]{yang2018implication}%
  \BibitemOpen
  \bibfield  {author} {\bibinfo {author} {\bibfnamefont {X.}~\bibnamefont
  {Yang}}\ and\ \bibinfo {author} {\bibfnamefont {M.}~\bibnamefont {Howland}},\
  }\bibfield  {title} {\bibinfo {title} {Implication of taylor's hypothesis on
  measuring flow modulation},\ }\href@noop {} {\bibfield  {journal} {\bibinfo
  {journal} {J. Fluid Mech.}\ }\textbf {\bibinfo {volume} {836}},\ \bibinfo
  {pages} {222} (\bibinfo {year} {2018})}\BibitemShut {NoStop}%
\bibitem [{\citenamefont {Wilczek}\ \emph {et~al.}(2015)\citenamefont
  {Wilczek}, \citenamefont {Stevens},\ and\ \citenamefont
  {Meneveau}}]{wilczek2015spatio}%
  \BibitemOpen
  \bibfield  {author} {\bibinfo {author} {\bibfnamefont {M.}~\bibnamefont
  {Wilczek}}, \bibinfo {author} {\bibfnamefont {R.~J.}\ \bibnamefont
  {Stevens}},\ and\ \bibinfo {author} {\bibfnamefont {C.}~\bibnamefont
  {Meneveau}},\ }\bibfield  {title} {\bibinfo {title} {Spatio-temporal spectra
  in the logarithmic layer of wall turbulence: large-eddy simulations and
  simple models},\ }\href@noop {} {\bibfield  {journal} {\bibinfo  {journal}
  {J. Fluid Mech.}\ }\textbf {\bibinfo {volume} {769}} (\bibinfo {year}
  {2015})}\BibitemShut {NoStop}%
\bibitem [{\citenamefont {Weber}\ \emph {et~al.}(2019)\citenamefont {Weber},
  \citenamefont {Reyers}, \citenamefont {Beck}, \citenamefont {Timme},
  \citenamefont {Pinto}, \citenamefont {Witthaut},\ and\ \citenamefont
  {Sch{\"a}fer}}]{weber2019wind}%
  \BibitemOpen
  \bibfield  {author} {\bibinfo {author} {\bibfnamefont {J.}~\bibnamefont
  {Weber}}, \bibinfo {author} {\bibfnamefont {M.}~\bibnamefont {Reyers}},
  \bibinfo {author} {\bibfnamefont {C.}~\bibnamefont {Beck}}, \bibinfo {author}
  {\bibfnamefont {M.}~\bibnamefont {Timme}}, \bibinfo {author} {\bibfnamefont
  {J.~G.}\ \bibnamefont {Pinto}}, \bibinfo {author} {\bibfnamefont
  {D.}~\bibnamefont {Witthaut}},\ and\ \bibinfo {author} {\bibfnamefont
  {B.}~\bibnamefont {Sch{\"a}fer}},\ }\bibfield  {title} {\bibinfo {title}
  {Wind power persistence characterized by superstatistics},\ }\href@noop {}
  {\bibfield  {journal} {\bibinfo  {journal} {Scientific reports}\ }\textbf
  {\bibinfo {volume} {9}},\ \bibinfo {pages} {1} (\bibinfo {year}
  {2019})}\BibitemShut {NoStop}%
\bibitem [{\citenamefont {Chowdhuri}\ \emph {et~al.}(2020)\citenamefont
  {Chowdhuri}, \citenamefont {Kalm{\'{a}}r-Nagy},\ and\ \citenamefont
  {Banerjee}}]{Chowdhuri_2020}%
  \BibitemOpen
  \bibfield  {author} {\bibinfo {author} {\bibfnamefont {S.}~\bibnamefont
  {Chowdhuri}}, \bibinfo {author} {\bibfnamefont {T.}~\bibnamefont
  {Kalm{\'{a}}r-Nagy}},\ and\ \bibinfo {author} {\bibfnamefont
  {T.}~\bibnamefont {Banerjee}},\ }\bibfield  {title} {\bibinfo {title}
  {Persistence analysis of velocity and temperature fluctuations in convective
  surface layer turbulence},\ }\href {https://doi.org/10.1063/5.0013911}
  {\bibfield  {journal} {\bibinfo  {journal} {Phys. Fluids}\ }\textbf {\bibinfo
  {volume} {32}},\ \bibinfo {pages} {076601} (\bibinfo {year}
  {2020})}\BibitemShut {NoStop}%
\bibitem [{\citenamefont {Laudani}\ \emph {et~al.}(2021)\citenamefont
  {Laudani}, \citenamefont {Zhang}, \citenamefont {Faouzi}, \citenamefont
  {Porcu}, \citenamefont {Ostoja-Starzewski},\ and\ \citenamefont
  {Chamorro}}]{Laudani_2021}%
  \BibitemOpen
  \bibfield  {author} {\bibinfo {author} {\bibfnamefont {R.}~\bibnamefont
  {Laudani}}, \bibinfo {author} {\bibfnamefont {D.}~\bibnamefont {Zhang}},
  \bibinfo {author} {\bibfnamefont {T.}~\bibnamefont {Faouzi}}, \bibinfo
  {author} {\bibfnamefont {E.}~\bibnamefont {Porcu}}, \bibinfo {author}
  {\bibfnamefont {M.}~\bibnamefont {Ostoja-Starzewski}},\ and\ \bibinfo
  {author} {\bibfnamefont {L.~P.}\ \bibnamefont {Chamorro}},\ }\bibfield
  {title} {\bibinfo {title} {On streamwise velocity spectra models with fractal
  and long-memory effects},\ }\href {https://doi.org/10.1063/5.0040453}
  {\bibfield  {journal} {\bibinfo  {journal} {Phys. Fluids}\ }\textbf {\bibinfo
  {volume} {33}},\ \bibinfo {pages} {035116} (\bibinfo {year}
  {2021})}\BibitemShut {NoStop}%
\end{thebibliography}%

\appendix
\section{Characteristic functional of the velocity field}
A first step towards a multipoint statistical description of a random field is the so-called fine-grained $n$-point-probability density function (PDF)
\begin{equation}
   \hat f_n(\mathbf{u}_1,\mathbf{x}_1;\ldots;\mathbf{u}_n,\mathbf{x}_n) =
  \prod_{i=1}^n \delta(\mathbf{u}_i-\mathbf{u}(\mathbf{x}_i))\;,
  \label{eq:fine-n-point}
\end{equation}
where $\mathbf{u}_i$ denote sample space variables
and the Dirac delta distributions guarantee that the fined-grained PDF is peaked at $\mathbf{u}_i=\mathbf{u}(\mathbf{x}_i)$ for each point $\mathbf{x}_i$.

The joint
$n$-point probability density function is obtained by averaging over different realizations of the random field $\mathbf{u}(\mathbf{x})$ according to
\begin{align}\nonumber
  f_n(\mathbf{u}_1,\mathbf{x}_1;\ldots;\mathbf{u}_n,\mathbf{x}_n) =& \left \langle  \hat f_n(\mathbf{u}_1,\mathbf{x}_1;\ldots;\mathbf{u}_n,\mathbf{x}_n) \right \rangle\\
   =&\left \langle
  \prod_{i=1}^n \delta(\mathbf{u}_i-\mathbf{u}(\mathbf{x}_i)) \right \rangle\;.
  \label{eq:n-point}
\end{align}
For later convenience, we define the so-called characteristic functional
\begin{equation}
  \varphi[\boldsymbol{\eta}(\mathbf{x})]  = \left \langle e^{i \int \textrm{d}\mathbf{x}
  \boldsymbol{\eta}(\mathbf{x}) \cdot \mathbf{u}(\mathbf{x})} \right \rangle\;.
\end{equation}
The joint $n$-point PDF (\ref{eq:n_point_3d}), for instance, is
related to the characteristic functional
\begin{equation}
  \varphi[\boldsymbol{\eta}(\mathbf{x})] =  \int_0^\infty \textrm{d}\xi g(\xi) e^{-\frac{1}{2} \int \textrm{d}\mathbf{x} \int \textrm{d}\mathbf{x}' \eta_\alpha(\mathbf{x})
  C_{\xi,\alpha \beta}(\mathbf{x}-\mathbf{x}') \eta_\beta(\mathbf{x}')}\;,
  \label{eq:charac}
\end{equation}
which can be shown by using the relation
\begin{align}
  \lefteqn{f_n(\mathbf{u}_1,\mathbf{x}_1;\ldots;\mathbf{u}_n,\mathbf{x}_n)} \\
  \nonumber
  =& \prod_{j=1}^n \int  \frac{\textrm{d} \boldsymbol{\eta}_j}{(2\pi)^{3n}} e^{-i \sum_{i=1}^n \boldsymbol{\eta}_i \cdot \mathbf{u}_i}\varphi\left[\boldsymbol{\eta}(\mathbf{x})=\sum_{i=1}^n \boldsymbol{\eta}_i \delta(\mathbf{x}-\mathbf{x}_i)\right].
\end{align}
\section{Implications of the incompressibility of the velocity field}
\label{app:incompressibility}
The incompressibility condition for the velocity field implies that $\nabla \cdot \mathbf{u}(\mathbf{x})=0$. The velocity correlation tensor can be derived from the characteristic functional (\ref{eq:charac}) according to
\begin{align}\nonumber
  \lefteqn{C_{\alpha \beta}(\mathbf{r})=\langle u_\alpha(\mathbf{x}+\mathbf{r}) u_\beta(\mathbf{x}) \rangle}\\ \nonumber
  =& \left. \frac{\delta}{\delta i \eta_{\alpha}(\mathbf{x}+\mathbf{r})}\frac{\delta}{\delta i \eta_{\beta}(\mathbf{x})} \varphi[\boldsymbol{\eta}(\mathbf{x})] \right|_{\boldsymbol{\eta}=0}\\
  =& \int_0^\infty \textrm{d}\xi g(\xi)C_{\xi,\alpha \beta}(\mathbf{r})\;.
\end{align}
Using the incompressibility condition, we obtain
\begin{equation}
  \frac{\partial}{\partial r_\alpha}C_{\alpha \beta}(\mathbf{r})=\left
   \langle  \frac{\partial  u_\alpha(\mathbf{x}+\mathbf{r})}{\partial r_\alpha} u_\beta(\mathbf{x}) \right \rangle =0\;,
\end{equation}
where summation over the same index is implied.
Inserting the tensorial form (\ref{eq:corr_tensor}) for $C_{\xi,\alpha \beta}(\mathbf{r})$, we obtain
\begin{align}
\lefteqn{0=\int \textrm{d}\xi g(\xi)   \frac{\partial}{\partial r_\alpha}C_{\xi,\alpha \beta}(\mathbf{r})=}\\ \nonumber
& \int \textrm{d}\xi g(\xi) \left[ \frac{1}{2r}\frac{\partial}{\partial r}r^2(C_{\xi,rr}(r)-
C_{\xi,tt}(r)) + \frac{\partial C_{\xi,tt}(r)}{\partial r}   \right] \frac{r_\beta}{r}\;,
\end{align}
which corresponds to Eq. (\ref{eq:karman}).
\section{Determination of longitudinal and transverse velocity increment statistics}
\label{app:longi}
First, we derive the longitudinal structure functions
\begin{equation}
  S_{n,r}(r) = \left \langle \left(  \left[ \mathbf{u}(\mathbf{x}+\mathbf{r})-\mathbf{u}(\mathbf{x})\right]\cdot \frac{\mathbf{r}}{r}\right)^n \right \rangle\;,
\end{equation}
for even $n$ (all moments of odd order are zero)
from the characteristic functional (\ref{eq:charac}). We thus obtain
\begin{align}\nonumber
  \lefteqn{S_{n,r}(r)= \left. \left(\left[\frac{\delta}{\delta i \boldsymbol{\eta}(\mathbf{x}+\mathbf{r})}-\frac{\delta}{\delta i \boldsymbol{\eta}(\mathbf{x})} \right]\cdot \frac{\mathbf{r}}{r} \right)^n  \varphi[\boldsymbol{\eta}(\mathbf{x})] \right|_{\boldsymbol{\eta}=0}}\\
  =&  \int_0^\infty \textrm{d}\xi g(\xi)(n-1)!! 2^{\frac{n}{2}} \left[C_{\xi,rr}(0)-C_{\xi,rr}(r)\right]^{\frac{n}{2}}\;.
\end{align}
On the other hand, the transverse structure functions
\begin{equation}
  S_{n,t}(r) = \left \langle \left(\frac{\mathbf{r}}{r} \times \left(\frac{\mathbf{r}}{r} \times \left[\mathbf{u}(\mathbf{x}+\mathbf{r})-\mathbf{u}(\mathbf{x}) \right] \right)\right)^n \right \rangle\;,
\end{equation}
for even $n$, can be derived from the characteristic functional (\ref{eq:charac}) according to
\begin{align} \nonumber
  \lefteqn{S_{n,t}(r)}\\ \nonumber
  =&\left. \left(\frac{\mathbf{r}}{r} \times  \left( \frac{\mathbf{r}}{r}\times \left[\frac{\delta}{\delta i \boldsymbol{\eta}(\mathbf{x}+\mathbf{r})}-\frac{\delta}{\delta i \boldsymbol{\eta}(\mathbf{x})} \right] \right) \right)^n  \varphi[\boldsymbol{\eta}(\mathbf{x})] \right|_{\boldsymbol{\eta}=0}\\
  =&  \int_0^\infty \textrm{d}\xi g(\xi)(n-1)!! 2^{\frac{n}{2}} \left[C_{\xi,tt}(0)-C_{\xi,tt}(r)\right]^{\frac{n}{2}}\;.
\end{align}
\section{Modification of the joint multipoint statistics due to the measurement points}
\label{app:modified}
For the case of a bridge scale mixture, the $n$-point statistics (\ref{eq:n_point_3d}) gets modified due to the correlation functions,
\begin{widetext}
\begin{align}
C_{\xi,\alpha \beta}^B(\mathbf{x},\mathbf{x}') = C_{\xi,\alpha \beta}(\mathbf{x}-\mathbf{x}')
  -  C_{\xi,\alpha \gamma}(\mathbf{x}-\mathbf{x}_i)  \left[\sigma_{\xi,i\gamma;j\delta}^{-1}-
  \sigma_{\xi,i\gamma;k\epsilon}^{-1} U_{\epsilon,k}
  U_{\zeta,l}  \sigma_{\xi,j\delta;l \zeta}^{-1} \right]
  C_{\xi,\delta \beta}(\mathbf{x}'-\mathbf{x}_j)
\end{align}
\end{widetext}
This expression now replaces Eq. (\ref{eq:corr_tensor})
and modifies the $n$-point statistics (\ref{eq:n_point_3d}). However, if $\mathbf{x}-\mathbf{x}' \ll |\mathbf{x}_i-\mathbf{x}_j|$, we can approximate $C_{\xi,\alpha \beta}^B(\mathbf{x},\mathbf{x}') \approx C_{\xi,\alpha \beta}(\mathbf{x}-\mathbf{x}')$.
Hence the small-scale statistics still coincide with the K62-type statistics.

\section{Sampling algorithm for the superstatistical random field}
\label{app:sample}
Here, we give a brief schematic depiction of the sampling algorithm discussed in Sec.~\ref{sec:sample}. In order to simplify the
presentation of the scheme, we simply consider a single component of the velocity field $u(x)$.
\begin{widetext}
\begin{algorithm}
  \textbf{Initialization of the scale mixture:}\\
  Assemble the noise vector $\hat \varphi(\mathbf{k})$
  in Fourier space\\
  \For{$i = 0;\ \KwTo\;  N_{\xi};$}{
  Draw a random number $\xi_i$ from the lognormal distribution
  $g(\xi)$\\
  Assemble the correlation function $C_{\xi_i}(\mathbf{r})$ and calculate its Fourier transform $\hat C_{\xi_i}(\mathbf{k})$\\
  Calculate the power spectrum $E_{\xi_i}(k)= \frac{1}{2\pi k^2} \hat C_{\xi_i}(\mathbf{k})$ \\
  Multiply the noise vector $\hat \varphi(\mathbf{k})$ by the amplitudes $E_{\xi_i}^{\frac{1}{2}}(k)$ which results in $\hat u(\xi_i,\mathbf{k})=E_{\xi_i}^{\frac{1}2}(k)\hat \varphi(\mathbf{k})$\\
  Inverse Fourier transform $\hat u(\xi_i,\mathbf{k})$ to obtain $u(\xi_i,\mathbf{x})$\\
  Optionally: Perform the bridge construction (\ref{eq:corr_lambda}) by linear operation on $u_\alpha(\xi_i,\mathbf{x})$, which transforms $u(\xi_i,\mathbf{x}) \rightarrow u^B(\xi_i,\mathbf{x})$
  }~\\
  ~\\
  \textbf{Perform the scale mixture:}\\
  \For{$r=|\mathbf{x}^2|;\ \KwTo\;  R;$}{
  Assign velocity field points ${u}(\xi,\mathbf{x})$ on spherical shell $r$ the same chosen parameter $\xi_i$ where $i$ is chosen randomly from $[0,N_\xi]$ \\
 Here, $i$ is chosen by an ordinary (integer) Ornstein-Uhlenbeck process in order to avoid the presence of fractional Gaussian noise in $u(\mathbf{x})$  \\
  $u(\mathbf{x})=u(\xi_i,\mathbf{x})$\\
  }
  \caption{Scheme for sampling a realization of the random field $u(\mathbf{x})$ from the joint multipoint PDF (\ref{eq:n_point_3d}). The optional point in the for-loop can be applied for constraining the random field on the measurement points $\mathbf{U}_i$ at points $\mathbf{x}_i$ by the bridge construction (\ref{eq:bridge}).}
  \label{algo}
\end{algorithm}
\end{widetext}
\end{document}